\documentclass[twocolumn]{aastex631}

\usepackage{graphicx}	
\usepackage{amsmath}	
\usepackage[version=4]{mhchem}
\usepackage{gensymb}
\usepackage{chemfig}
\usepackage{makecell}
\usepackage{tabularx}
\usepackage{booktabs}

\newcommand\numberthis{\addtocounter{equation}{1}\tag{\theequation}}

\newcolumntype{b}{X}
\newcolumntype{s}{>{\hsize=.5\hsize}X}

\received{2024 Jul 12}
\revised{2024 Sep 23}
\accepted{2024 Oct 20}

\submitjournal{ApJ}

\shorttitle{PHO exoplanet photochemistry}
\shortauthors{Lee et al.}
\graphicspath{{./}{plots/}}

\begin{document}

\title{A photochemical PHO network for hydrogen-dominated exoplanet atmospheres}

\correspondingauthor{Elspeth K.H. Lee}
\email{elspeth.lee@unibe.ch}

\author[0000-0002-3052-7116]{Elspeth K.H. Lee}
\affiliation{Center for Space and Habitability, University of Bern, Gesellschaftsstrasse 6, CH-3012 Bern, Switzerland}

\author[0000-0002-8163-4608]{Shang-Min Tsai}
\affiliation{Department of Earth Sciences, University of California, Riverside, CA 92521, USA}

\author[0000-0002-8837-0035]{Julianne I. Moses}
\affiliation{Space Science Institute, Boulder, CO 80301, USA}

\author[0000-0003-3648-6893]{John M.C. Plane}
\affiliation{School of Chemistry, University of Leeds, Leeds LS2 9JT, UK}

\author[0000-0001-6627-6067]{Channon Visscher}
\affiliation{Department of Chemistry, Dordt University, Sioux Center, IA 51250, USA}
\affiliation{Space Science Institute, Boulder, CO 80301, USA}

\author[0000-0001-6297-9187]{Stephen J. Klippenstein}
\affiliation{Chemical Sciences and Engineering Division, Argonne National Laboratory, Lemont, IL 60439, USA}

\begin{abstract}
Due to the detection of phosphine (\ce{PH3}) in the Solar System gas giants Jupiter and Saturn, \ce{PH3} has long been suggested to be detectable in exosolar substellar atmospheres too.
However, to date, a direct detection of phosphine has proven to be elusive in exoplanet atmosphere surveys.
We construct an updated phosphorus-hydrogen-oxygen (PHO) photochemical network suitable for simulation of gas giant hydrogen-dominated atmospheres.
Using this network, we examine PHO photochemistry in hot Jupiter and warm Neptune exoplanet atmospheres at Solar and enriched metallicities.
Our results show for HD 189733b-like hot Jupiters that HOPO, PO and \ce{P2} are typically the dominant P carriers at pressures important for transit and emission spectra, rather than \ce{PH3}.
For GJ1214b-like warm Neptune atmospheres our results suggest that at Solar metallicity \ce{PH3} is dominant in the absence of photochemistry, but is generally not in high abundance for all other chemical environments.
At 10 and 100 times Solar, small oxygenated phosphorus molecules such as HOPO and PO dominate for both thermochemical and photochemical simulations.
The network is able to reproduce well the observed \ce{PH3} abundances on Jupiter and Saturn.
Despite progress in improving the accuracy of the PHO network, large portions of the reaction rate data remain with approximate, uncertain or missing values, which could change the conclusions of the current study significantly.
Improving understanding of the kinetics of phosphorus-bearing chemical reactions will be a key undertaking for astronomers aiming to detect phosphine and other phosphorus species in both rocky and gaseous exoplanetary atmospheres in the near future.
\end{abstract}

\keywords{Exoplanet atmospheres(487) -- Exoplanet atmospheric composition (2021) -- Chemical kinetics(2233)}

\section{Introduction} 
\label{sec:intro}

Phosphine (PH3) has been detected in the gas giants of the Solar System since the 1970's, in Jupiter \citep{Ridgway1976,Ridgway1976b} and Saturn \citep{Bregman1975}, providing strong evidence of non-equilibrium chemistry and vertical mixing occurring in their atmospheres, as \ce{P4O6} was the expected dominant P carrier at Jupiter/Saturn's photospheric temperatures at chemical equilibrium \citep[e.g.][]{Fegley1985,Fegley1994,Visscher2005}.

Initial chemical kinetic modelling of phosphorous species in the atmospheres of Jupiter and Saturn began with \citet{Prinn1975}, using
available experimental data at the time and investigated the formation pathways of \ce{P4} and solid \ce{P4}(s) from the initial photodisoociation of \ce{PH3}.
Later models focused on the coupled photochemistry of \ce{PH3} and \ce{NH3} and the likelihood of diphosphine (\ce{P2H4}) as the primary phosphorus-containing tropospheric aerosol on the Jovian planets \citep{Strobel1977, Kaye1983, Kaye1984, Edgington1998}.
More recent studies of the non-equilibrium abundance of \ce{PH3} on Jupiter and Saturn emphasize different quench pathways, rate-limiting reactions, and important P-bearing species \citep{Wang2016}.

Outward from the context of Solar System gas-giants, phosphine was also expected to be present and detectable in brown dwarf and hydrogen-dominated exoplanet atmospheres \citep{Fegley1996,Visscher2006}.
However, to date, direct detection of \ce{PH3} in brown dwarf atmospheres has proven elusive.
Several studies where \ce{PH3} was expected to be detected in the brown dwarf regime failed to detect signatures of \ce{PH3} absorption \citep[e.g.][]{Morley2018, Miles2020, Beiler2023}.
With the advent of the JWST telescope, which can now distinguish signatures of trace gas phase species such as \ce{SO2} in warm Saturn atmospheres \citep{Rustamkulov2023,Alderson2023,Powell2024}, there is an opportunity to detect \ce{PH3} signatures and other trace phosphorus-bearing species in exoplanet atmospheres with more clarity.

Phosphine has long been suggested to be a biomarker indicator due to the importance of phosphorus to the functioning and development of Earth-based biological organisms \citep[e.g.][]{Sousa-Silva2020}.
Phosphorus has a wide range of redox states \citep{Pasek2017} and the accumulation of phosphate seems to be key to the origin of life \citep{Toner2019}. 
Simulations performed in \citet{Angerhausen2023} suggest that the proposed ESA LIFE mission \citep{Quanz2022} will be able to detect \ce{PH3} in cold super-Earth and Jupiter-like exoplanets in under one hour of observing time.
For smaller planets, their simulations suggest ten hours of observing time to detect \ce{PH3}. 

Recently, hints for \ce{PH3} production in Venus's atmosphere were seen in microwave measurements \citep{Greaves2021}\footnote{Though not without controversy \citep[e.g.][]{Encrenaz2020,Villanueva2021}.}, which could not be explained by non-biological kinetic modelling alone \citep{Bains2021}.
If \ce{PH3} production occurs, it suggests that active biology or unknown abiotic chemistry may be occurring in the upper atmosphere of Venus. 
However, full confirmation and an accurate determination of \ce{PH3} abundance and vertical profile \citep{Lincowski2021} on Venus may need to wait for proposed Venus orbiter and probe missions \citep[e.g.][]{Ghail2017, Garvin2022} and other dedicated search efforts.

In summary, from the above studies, \ce{PH3} is a key molecule to explore and search for across the planetary parameter regime. 
From large gas giants to small rocky planets, understanding the formation chemistry of \ce{PH3} and other P-bearing molecules will be a significant goal for astronomers in the near and long term.

To start to meet this challenge, in this study, we develop and investigate the properties of a phosphorus-hydrogen-oxygen (PHO) photochemical network suitable for hydrogen-dominated atmospheres.
We aim to elucidate the mechanisms at play that give rise to \ce{PH3} and other phosphorus species in exoplanet atmospheres and study the role of photochemistry in the PHO system.
In Section \ref{sec:network}, we provide details on the thermochemical, kinetic and photochemical aspects of the proposed network.
In Section \ref{sec:comp_wang}, we compare the results of our thermochemical scheme to those presented in \citet{Wang2017}.
In Section \ref{sec:hj}, we apply our scheme to a canonical hot Jupiter atmosphere, examining the effects of photochemical processes on the P species in these atmospheres.
In Section \ref{sec:wn}, we move to colder and small warm Neptune planets, examining the phosphorus content of their atmospheres, in particular the effects of metallicity on the dominant P species carriers.
Section \ref{sec:djs} applies the network to the deep atmospheres of Jupiter and Saturn.
Section \ref{sec:disc} contains a discussion of the results and the mechanisms and Section \ref{sec:obs_cons} examines the potential observational impacts.
Section \ref{sec:conc} summarises the conclusions of the study.

\section{Development of an exoplanet PHO photochemical network}
\label{sec:network}

To perform the kinetic modelling of the PHO network, we use the 1D photochemical model VULCAN \citep{Tsai2017,Tsai2021} to integrate the network to a steady state.
Throughout, we assume the Solar elemental ratios (or some multiple factor there of) from \citet{Asplund2021} for each element.
Chemical equilibrium is assumed for each species as their initial conditions, which is performed using the FastChem \citep{Stock2018} module of VULCAN.
The PHO photochemical scheme and related data can be found as part of the publicly available VULCAN\footnote{\url{https://github.com/exoclime/VULCAN}} code.

\subsection{PHO thermochemical kinetics}
\label{sec:thermo}

The basis for the thermochemistry scheme comes from the network of \citet{Wang2016}, originally designed to investigate \ce{PH3} in the deep Jupiter and Saturn atmospheres and subsequently applied to hot Jupiter atmospheres in \citet{Wang2017}. 
This primarily pulled data from the network of \citet{Twarowski1995}, developed for flame and ignition applications.
However, \citet{Twarowski1995} used Benson group-additivity rules \citep{Benson1958} to estimate the activation energies, and also estimated the rate constants of the majority of their reaction list, therefore making the network highly approximate overall.
Despite its approximate nature, the \citet{Twarowski1995} network provides a useful basis for the construction of a phosphorus photochemical network suitable for exoplanet atmospheres and give indications of the important chemical pathways that are required to be studied in more detail.
For the phosphorous reactions, we take the reaction list used in \citet{Wang2016}\footnote{Which can be found on KIDA: \url{https://kida.astrochem-tools.org/}} as an initial starting point for the PHO thermochemical network.

For the HO chemistry, we use the species and reactions from \citet{Tsai2017, Tsai2021} (Appendix \ref{app:list}). 
In addition, we include photolysis reactions for \ce{H2O}, \ce{H2}, OH, \ce{HO2} and \ce{O2} (Table, \ref{tab:phchem}).

Since \citet{Twarowski1995}, several studies have attempted to improve the accuracy of key reaction rates through various experimental and theoretical efforts.
\citet{Haworth2002} and \citet{Mackie2002} investigated several uncertain phosphorus oxidation reactions using computational chemistry techniques and updated their rates.
\citet{Jayaweera2005} updated several reactions from \citet{Twarowski1995} with theoretically derived rates and estimations, mostly stemming from the results presented in \citet{Glaude2000}.
Several phosphorus oxidation reactions were also investigated experimentally by \citet{Douglas2019} and \citet{Douglas2020}.
These new rates were subsequently applied in stellar wind modelling \citep{Douglas2022} and the modelling of P chemistry in the Earth's upper atmosphere, where P is produced by the ablation of cosmic dust particles during atmospheric entry \citep{Plane2021}.
\citet{Baptista2023} calculated several high pressure rates for PH$_{x}$ decomposition reactions. 

In addition to incorporating updated rates from the above studies, we have also produced new theoretical rate coefficients for some key reactions:
\begin{align}
    \ce{H} + \ce{PH3} &\rightarrow \ce{H2} + \ce{PH2} \tag{R55} \\
    \ce{H} + \ce{PH2} &\rightarrow \ce{H2} + \ce{PH}  \tag{R57} \\
   \ce{H} + \ce{PH} &\rightarrow \ce{P} + \ce{H2}  \tag{R59} \\
    \ce{P} + \ce{PH} &\rightarrow \ce{H} + \ce{P2} \tag{R281} \\
    \ce{PH2} + \ce{PH} &\rightarrow \ce{P} + \ce{PH3} \tag{R293} \\
    \ce{H} + \ce{PH2} + \ce{M} &\rightarrow \ce{PH3} + \ce{M} \tag{R329} \\
    \ce{H} + \ce{P2} + \ce{M} &\rightarrow \ce{P2H} + \ce{M} \tag{R335} \\
    \ce{PO} + \ce{PO2} + \ce{M} &\rightarrow \ce{P2O3} + \ce{M}  \tag{R337} \\
    \ce{H} + \ce{PH} + \ce{M} &\rightarrow \ce{PH2} + \ce{M} \tag{R341} \\
    \ce{P2O3} + \ce{P2O3} + \ce{M} &\rightarrow \ce{P4O6} + \ce{M}  \tag{R349} \nonumber
\end{align}
Notably, we have included a theoretical reaction rate to form \ce{P4O6} from the recombination reaction R349, which is detailed in Appendix \ref{app:P4O6}.
Furthermore, we have also included estimates for the reaction rates involving the formation of \ce{P2H2} and \ce{P2H4} based on their nitrogen counterparts (Appendix \ref{app:list}).
Overall, we have devised a PHO network that replaces around 25$\%$ of the original \citet{Twarowski1995}/\citet{Wang2016} network, amounting to a total of 32 species with 195 forward reactions (390 total including reverse reactions) plus 18 irreversible photochemical reactions.
Appendix \ref{app:list} presents the list reactions in the PHO network and their rate coefficients.

\subsubsection{Phosphorus thermochemical data}

A large area of uncertainty in chemical modelling of phosphorus species is the accuracy of available thermochemical data and choice of database.
Of note are the different values adopted for the \ce{P4O6} enthalpy of formation, as using different sources changes the expected equilibrium distribution of P-bearing species at cooler temperatures \citep[see discussion in][]{Fegley1994, Borunov1995, Wang2016, Visscher2020}.
For example, at temperatures relevant for Jupiter's and Saturn's deep atmosphere, the expected dominant P-bearing gas at equilibrium can be \ce{P4O6} (using \ce{P4O6} enthalpy values from NIST-JANAF; \citealt{Chase1998}) or \ce{H3PO4} (using \ce{P4O6} enthalpy values from \citealt{Gurvich1989}).

\citet{Wang2016} also discuss this discrepancy, opting for the thermodynamic data from \citet{Gurvich1989} as incorporated into the NASA thermodynamic polynomials \citep[e.g.,][]{McBride1992,Zehe2002} which favors the formation of \ce{H3PO4} at low temperatures.
Recently, the \citet{Bains2023} review of \ce{P4O6} thermochemistry suggests that the commonly used NIST-JANAF database \citep{Chase1998} values for the free energy of formation of \ce{P4O6} are likely too low and the molecule is less stable than the NIST-JANAF values suggest. 
In addition, \citet{Lodders1999} update the thermochemical properties of PH, \ce{PH3} and PN with the white phosphorus reference state which were not corrected in the NIST-JANAF database \citep{Chase1998}.

In the present work, we likewise adopt thermodynamic values from the NASA database \citep[including \ce{P4O6} enthalpy data from the Gurvich database;][]{McBride1992, Zehe2002} and \citet{Burcat2005} for simplicity and consistency between \citet{Wang2017} and our study. 
These values are also used in the FastChem \citep{Stock2018} module to VULCAN, which calculates the initial conditions of each species in chemical equilibrium.

\subsection{PHO Photochemistry}
\label{sec:photo}

\begin{figure}
    \centering
    \includegraphics[width=0.49\textwidth]{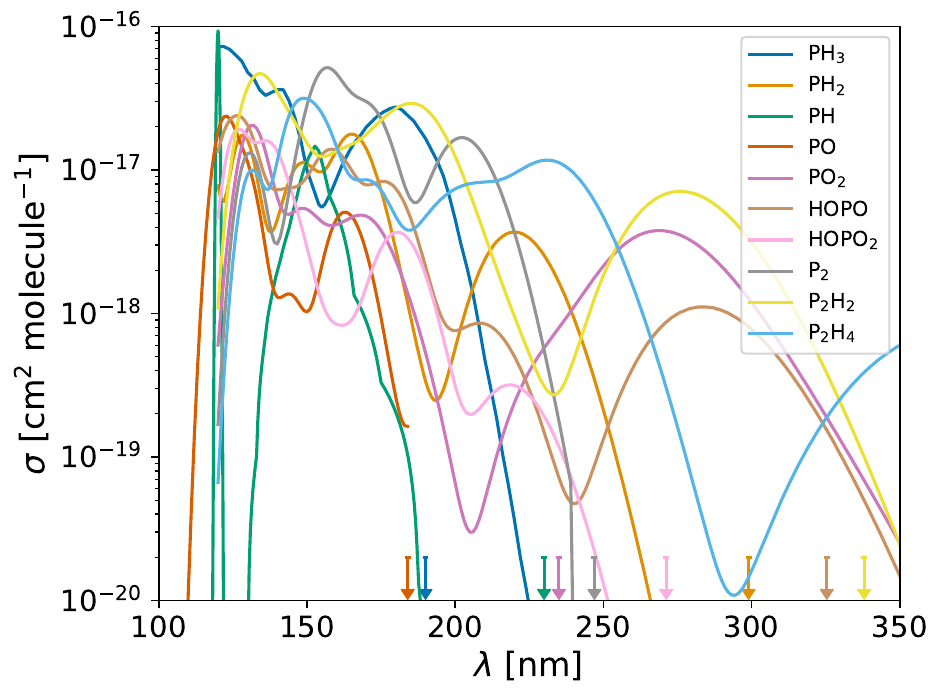}
    \caption{UV absorption cross sections for the phosphorus species in the network that undergo photolysis reactions.
    The photolysis threshold wavelength for each species is indicated by the coloured arrows.}
    \label{fig:P_UV}
\end{figure}

\begin{table*}
    \label{tab:phchem}
    \caption{List of photolysis reactions used for the PHO photochemical network.}
    \centering
    \begin{tabular}{c|l|c|c} \hline \hline
      Species & Reaction & Threshold (nm) & Cross section/branching ratio reference \\ \hline   
     \ce{H2O}  & $\rightarrow$ H + OH & 207 & \citet{Heays2017}\\
       & $\rightarrow$ \ce{H2} + O($^{1}$D) &  & \citet{Huebner2015} \\
       & $\rightarrow$ O + H + H & 145  & \\
     \ce{H2}  & $\rightarrow$ H + H & 120 &  \citet{Heays2017} \\
     OH  & $\rightarrow$ H + O & 265 &  \citet{Heays2017}  \\
     \ce{HO2}  & $\rightarrow$ O + OH & 275 &   \citet{Heays2017}\\
     \ce{O2}  & $\rightarrow$ O + O & 240 &   \citet{Huebner2015}\\    
      & $\rightarrow$ O + O($^{1}$D) & 175.6 &   \citet{Sander2006}\\   
     PH  & $\rightarrow$ P + H & 190 &  \citet{Heays2017}\\
     \ce{PH2}  & $\rightarrow$ PH + H & 299 &  This study \\
     \ce{PH3}  & $\rightarrow$ \ce{PH2} + H & 230   & \citet{Chen1991} \\   
    \ce{PO} & $\rightarrow$ P + O & 184  & This study \\ 
    \ce{PO2} & $\rightarrow$ PO + O & 235  & This study \\ 
     HOPO  & $\rightarrow$ \ce{PO2} + H & 325.4   & \citet{Plane2021} \\ 
     \ce{HOPO2} & $\rightarrow$ \ce{PO2} + OH & 271.2  & \citet{Plane2021} \\ 
     \ce{P2} & $\rightarrow$ P + P & 247  & This study \\ 
     \ce{P2H2} & $\rightarrow$ PH + PH & 338  & This study \\ 
     \ce{P2H4} & $\rightarrow$ \ce{PH2} + \ce{PH2} & 508  & This study \\ \hline
    \end{tabular} 
\end{table*}

For the PHO photochemical network, we include several photolysis reactions listed in Table \ref{tab:phchem}. 
We take UV cross sections from the PhiDrates \citep{Huebner2015} and Leiden Observatory \citep{Heays2017} databases, with the \ce{PH3} UV cross-sections taken from \citet{Chen1991}.
HOPO and \ce{HOPO2} cross-sections are taken from the theoretical calculations of \citet{Plane2021}.
We also calculate new theoretical UV cross-sections and threshold wavelengths for \ce{PH2}, PO, \ce{PO2}, \ce{P2}, \ce{P2H2} and \ce{P2H4} (Appendix \ref{app:P4O6}).
In Figure \ref{fig:P_UV}, we show the UV photo-absorption cross-sections and threshold wavelengths of each of the phosphorus species that undergo photolysis.
This expands the total number of photolysis reactions involving P to ten.

\section{Comparison to Wang et al. (2017)}
\label{sec:comp_wang}

\begin{figure*}
    \centering
    \includegraphics[width=0.49\textwidth]{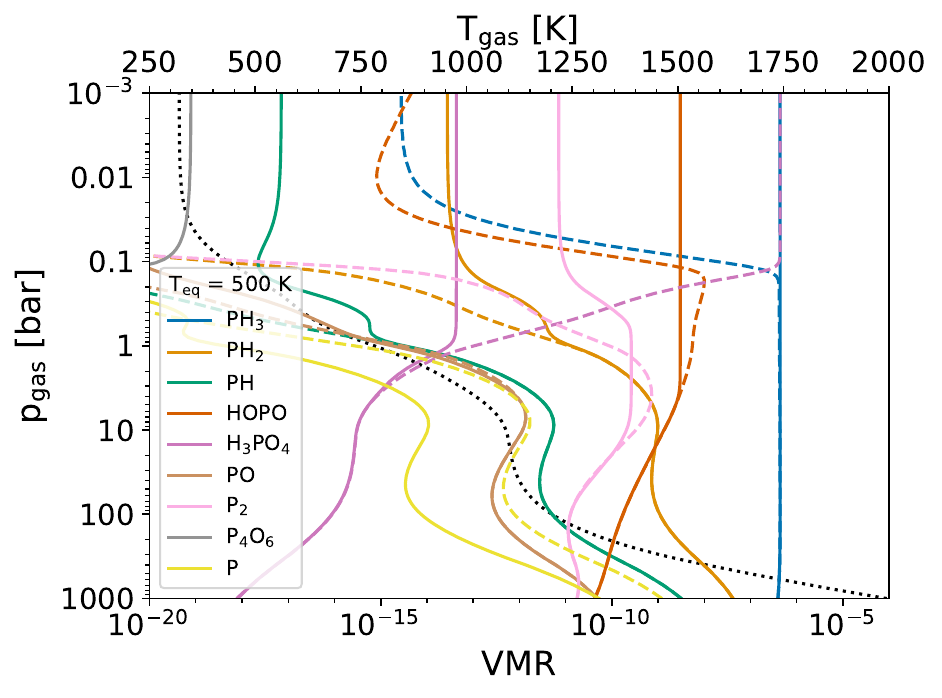}
    \includegraphics[width=0.49\textwidth]{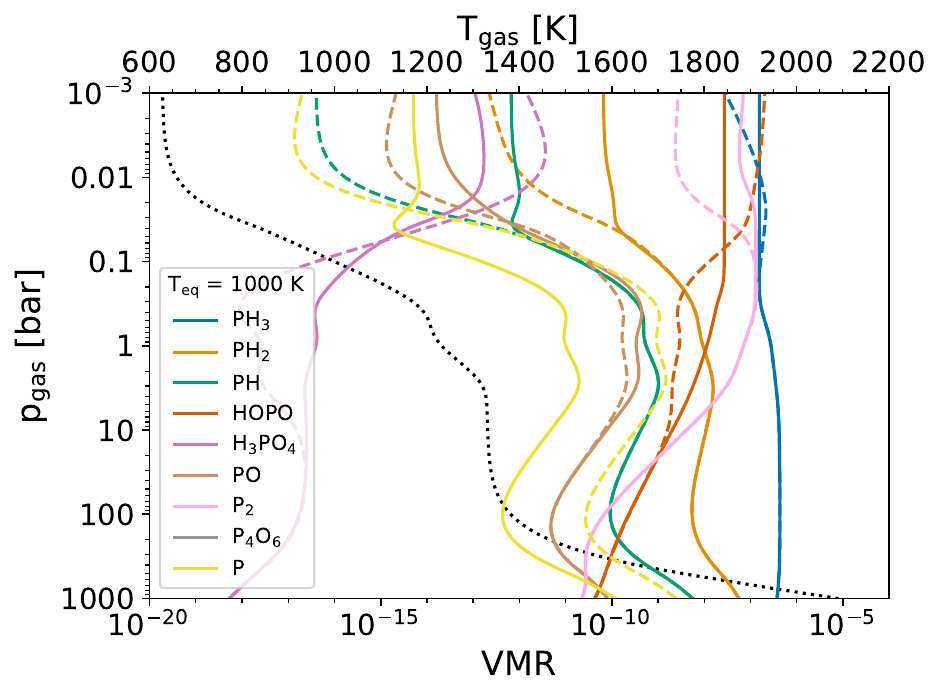}
    \includegraphics[width=0.49\textwidth]{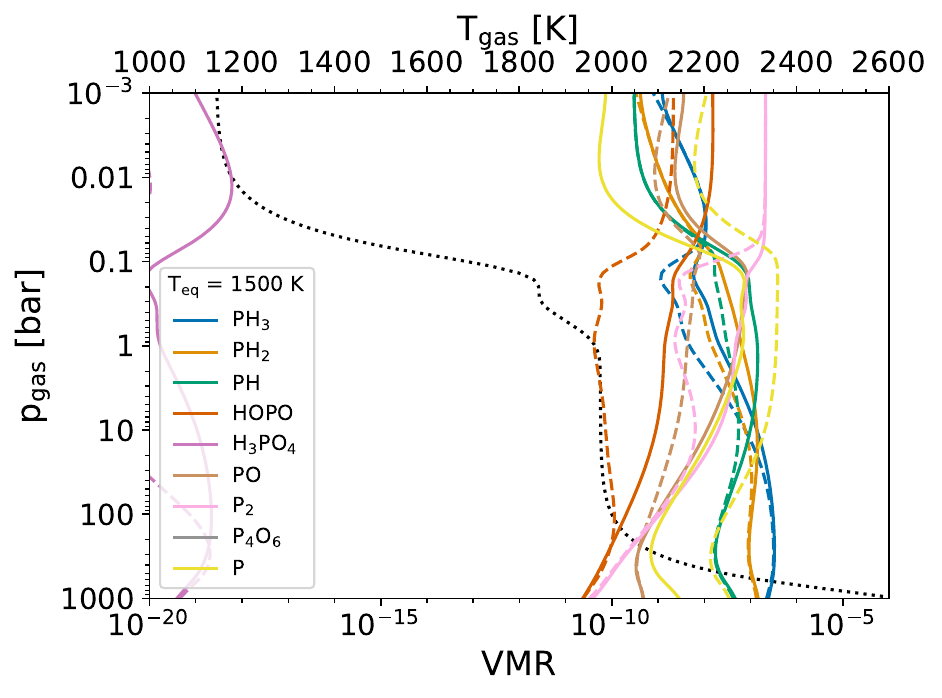}
    \includegraphics[width=0.49\textwidth]{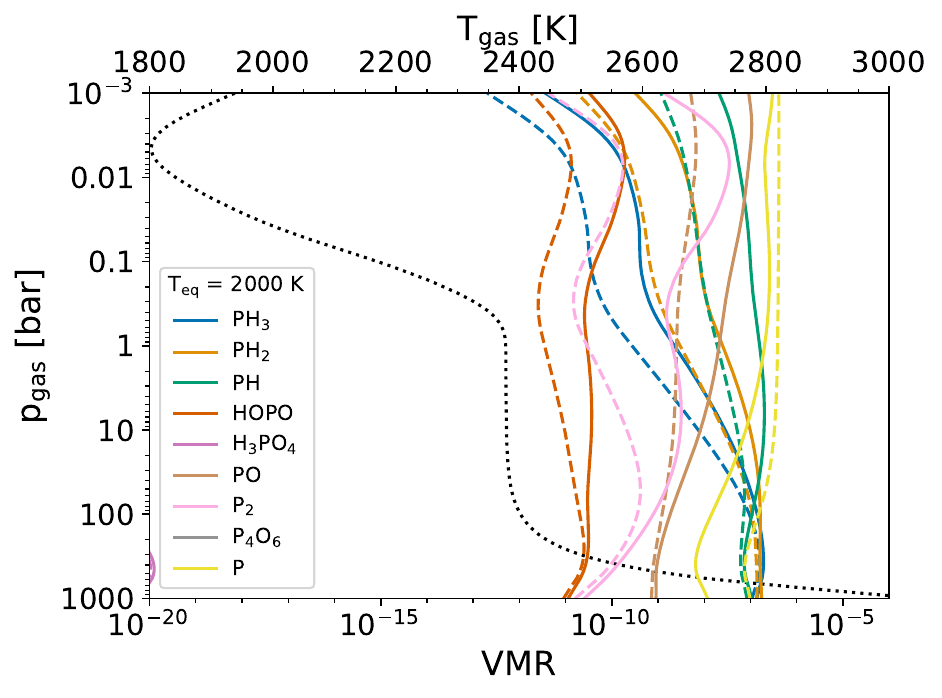}
    \caption{VULCAN PHO network results of volume mixing ratios (VMR; coloured solid lines) for comparison to the \citet{Wang2017} (their Figure 2.) results.
    The chemical equilibrium volume mixing ratios are denoted by the coloured dashed lines.
    This shows results for various PHO species of interest across the hot-Jupiter T-p profiles with equilibrium temperatures T$_{\rm eq}$ =  500, 1000, 1500 and 2000 K (black dotted lines) \citep{Wang2017}. 
    A constant K$_{\rm zz}$ = 10$^{9}$ cm$^{2}$ s$^{-1}$ and Solar metallicity is assumed as in \citet{Wang2017}.}
    \label{fig:Teq}
\end{figure*}

\begin{figure*}
    \centering
    \includegraphics[width=0.49\linewidth]{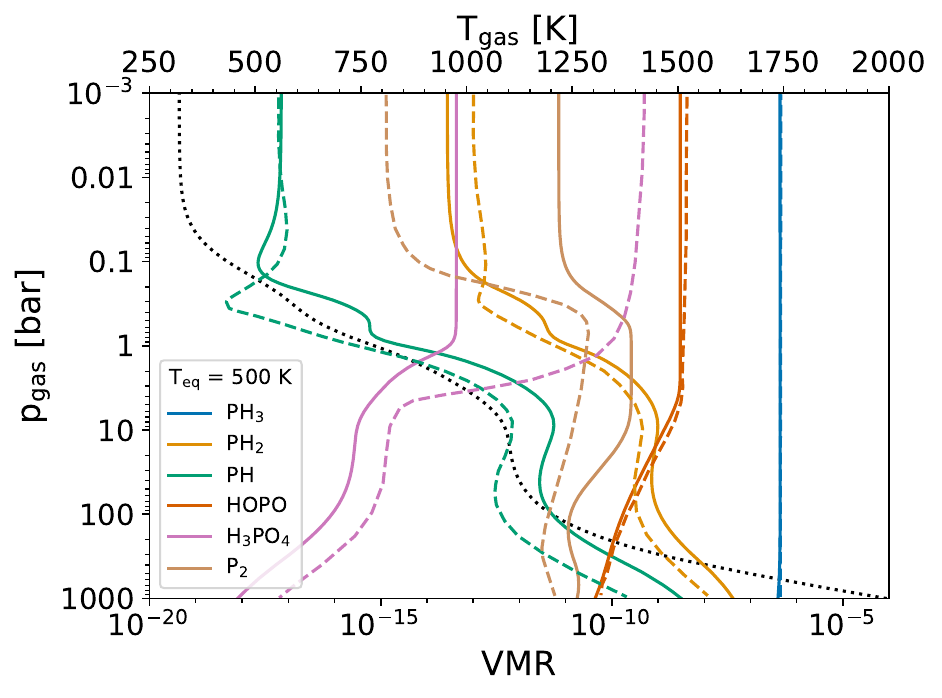}
    \caption{The updated PHO network (solid lines) compared to the results in \citet{Wang2017}(dashed lines) for the T$_{\rm eq}$ = 500 K test case from \citet{Wang2017}.}
    \label{fig:Wang_comp}
\end{figure*}

\begin{figure*}
    \centering
    \includegraphics[width=0.49\linewidth]{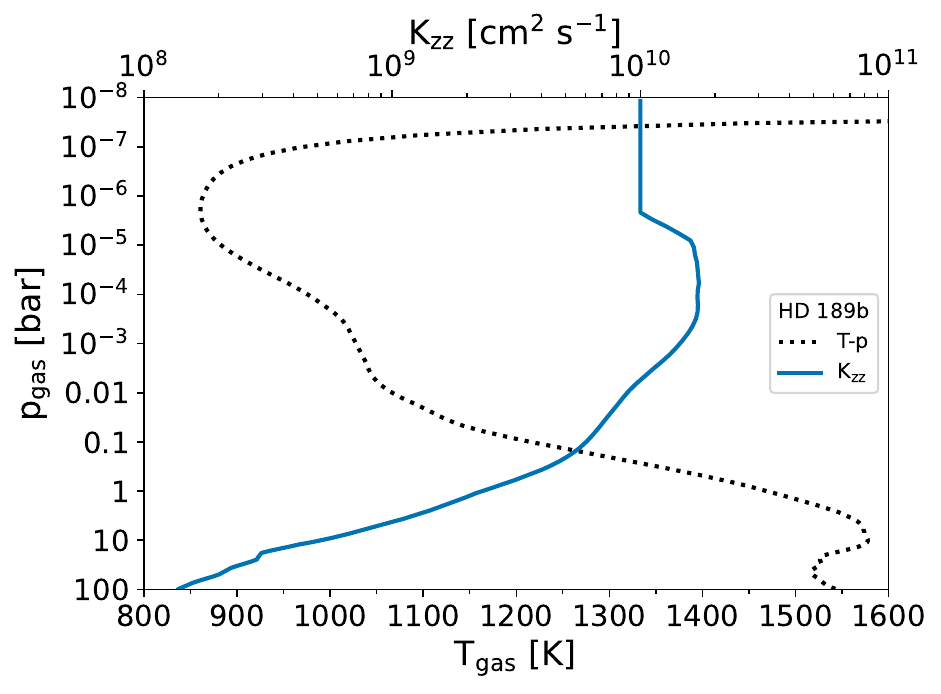}
    \includegraphics[width=0.49\linewidth]{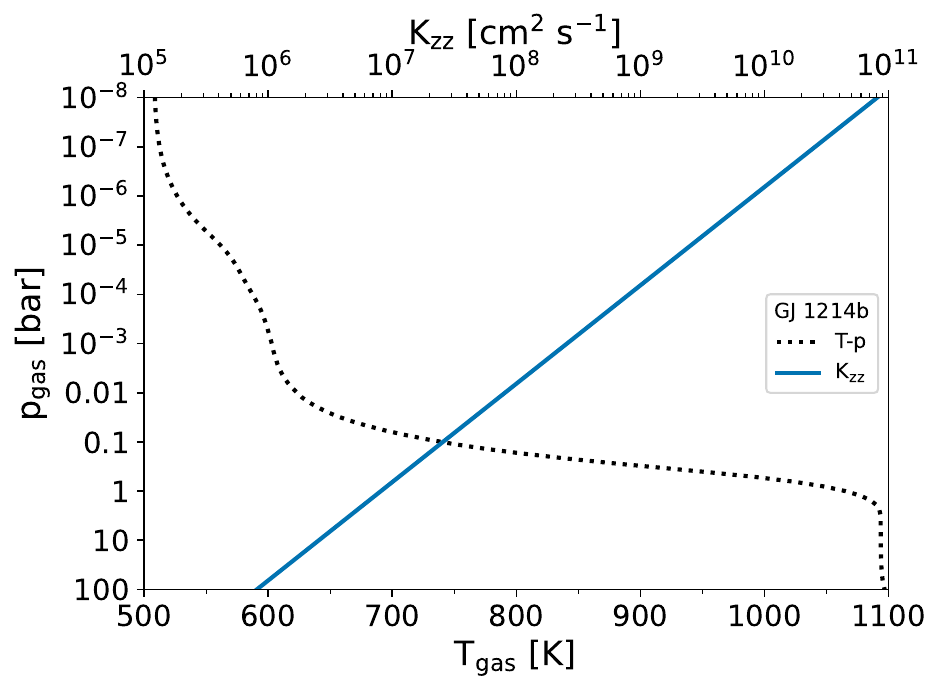}
    \caption{Input T-p and K$_{\rm zz}$ profiles for the HD 189733b case (left) and GJ 1214b case (right) used in this study.}
    \label{fig:T_p_Kzz}
\end{figure*}

In this section, we compare our thermochemical kinetics and transport scheme to that of \citet{Wang2017}, who applied the network of \citet{Wang2016} to various exoplanet temperature-pressure (T-p) profiles and vertical mixing rate scenarios.
In Figure \ref{fig:Teq}, we show the results of the T$_{\rm eq}$ = 500 K, 1000 K, 1500 K and 2000 K hot-Jupiter models that use the same T-p profiles as in \citet{Wang2017} and K$_{\rm zz}$ = 10$^{9}$ cm$^{2}$ s$^{-1}$ at Solar metallicity.
Our results agree well across all equilibrium temperatures, with the major difference being the abundance of \ce{H3PO4} in the 1500 and 2000 K profiles.
However, the volume mixing ratio (VMR) of \ce{H3PO4} is extremely small here and in \citet{Wang2017}, making it a very minor species at these higher temperatures.
Other minor differences are in the oxygenated phosphorus species such as HOPO, which we produce slightly less of.
We attribute this difference to the updated reaction rates used here, which results in a reduction in the number of oxygen radicals able to oxidise P, as well as the specific updated rates for the formation of HOPO and other oxygenated P molecules.

In Figure \ref{fig:Wang_comp}, we compare the T$_{\rm eq}$ = 500 K results directly from \citet{Wang2017} and the new network.
We produce consistent profiles for \ce{PH3} and HOPO, but differences are seen for the other molecules, suggesting the new network generally produces different results to the \citet{Wang2017} network.
We find larger differences in the upper atmosphere, in particular \ce{H3PO4} which is different by four orders of magnitude between the models.
\ce{P2} also shows large difference, by four order of magnitude in the upper atmosphere.
Overall, these results show a relative level of consistency between our study and \citet{Wang2017} at least for the major species \ce{PH3} and HOPO, but large differences are seen for the minor species.
This suggests that the updated rates in the new network significantly alters the chemical profiles compared to the \citet{Wang2017} network.

\section{Application to hot Jupiter atmospheres}
\label{sec:hj}

\begin{figure*}
    \centering
    \includegraphics[width=0.49\textwidth]{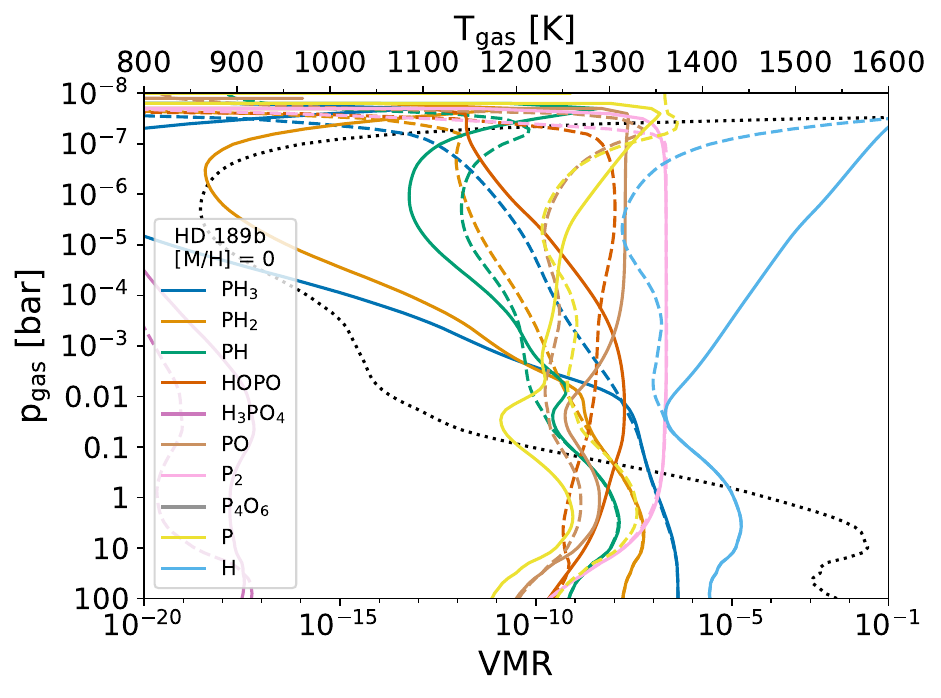}
    \includegraphics[width=0.49\textwidth]{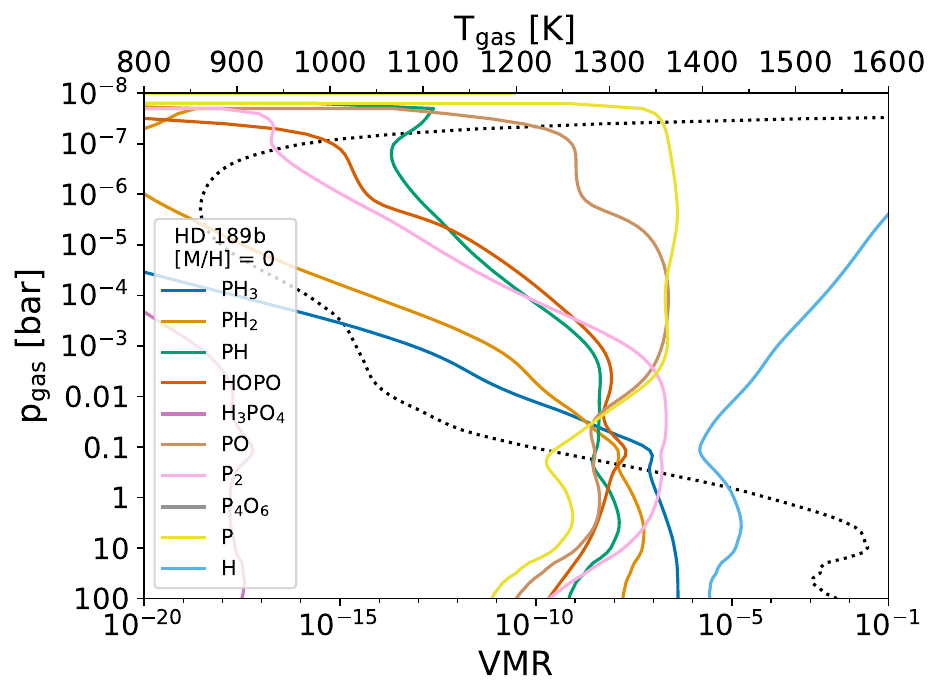}
    \includegraphics[width=0.49\textwidth]{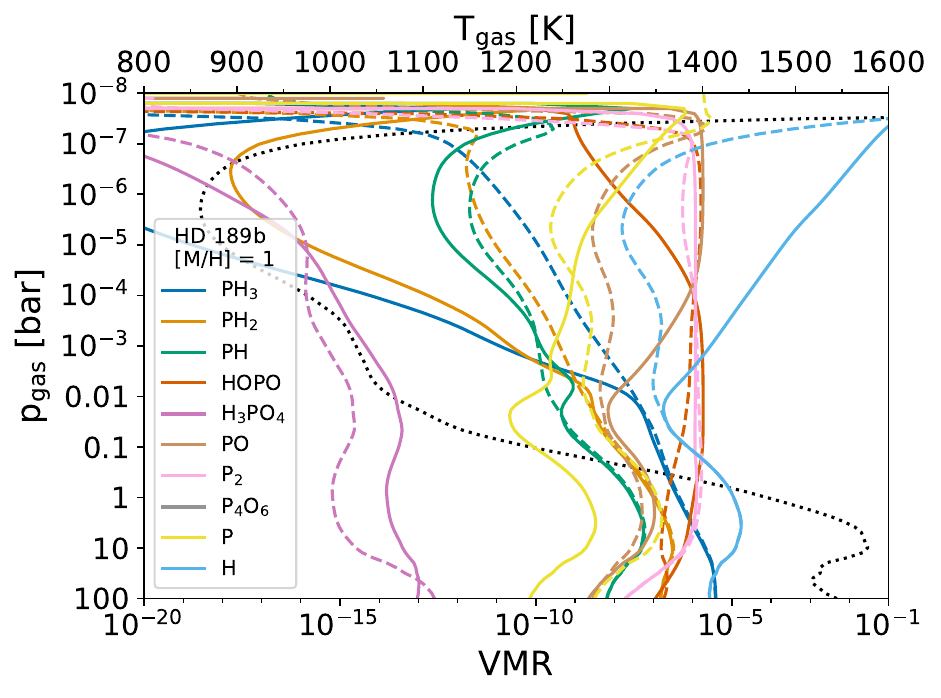}
    \includegraphics[width=0.49\textwidth]{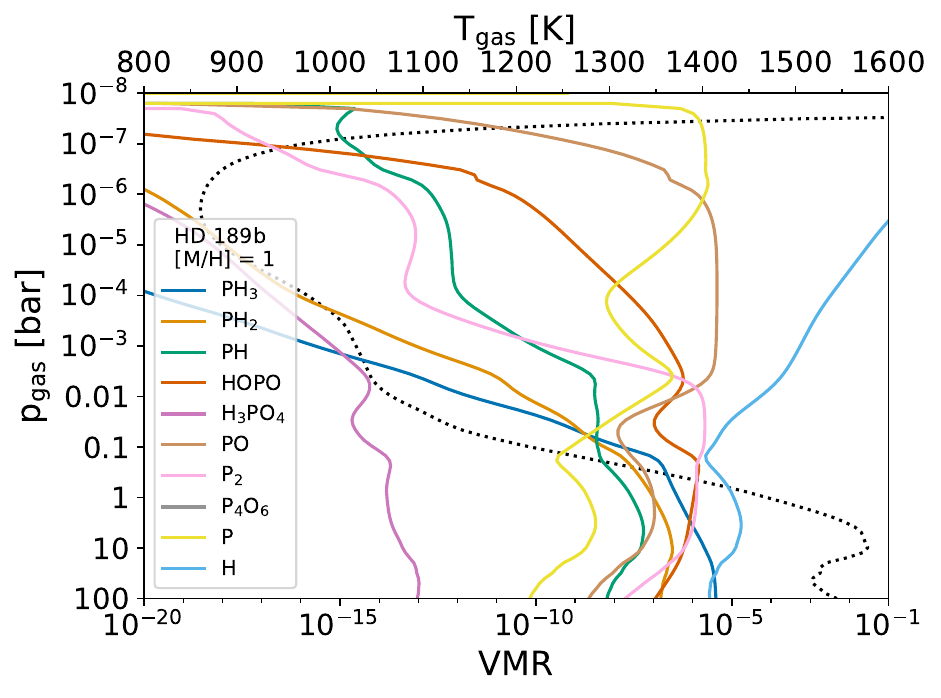}
    \caption{VULCAN PHO network results for the benchmark 1D HD 189733b case.
    The volume mixing ratio (VMR) of each species is shown as solid lines and the T-p profile as a black dotted line.
    The dashed lines denote the chemical equilibrium values.
    Top left: thermochemistry only at [M/H] = 0. 
    Top right: photochemical model at [M/H] = 0.
    Bottom left: thermochemistry only at [M/H] = 1.
    Bottom right: photochemical model at [M/H] = 1.
     }
    \label{fig:HD189}
\end{figure*}

In this section, we apply the PHO photochemical network to the benchmark hot Jupiter, HD 189733b 1D model parameters presented in \citet{Moses2011}, taking the T-p, K$_{\rm zz}$ and stellar flux model from that study.
We perform a Solar and 10 times Solar metallicity model for a thermochemistry only and photochemical test, and assess the impact of photochemistry on the vertical profiles of PHO species.
Figure \ref{fig:T_p_Kzz} shows the input T-p and K$_{\rm zz}$ profile.

Figure \ref{fig:HD189} presents the results of the model calculations.
In the thermochemical-kinetics only models, our results suggest that at Solar metallicity \ce{P2} is the main P carrier at pressure levels less than 1 bar, generally maintaining its equilibrium abundance throughout.
\ce{PH3} is generally confined to the deeper atmosphere ($>$ 1 bar), and at chemical equilibrium.
For 10 times Solar, the atmosphere becomes more oxygenated, with HOPO, PO and \ce{P2} dominating the atmospheric composition.
Our results indicate a rapid reaction pathway producing HOPO and PO, significantly pushing them and \ce{PH3} out of equilibrium in the upper atmosphere.
\ce{PH3} is now confined to the very deep atmosphere at pressures greater than 10 bar.

Comparing the T$_{\rm eq}$ = 1000 K from Figure \ref{fig:Teq} to our HD 189733b thermochemical kinetics only results shows a similar \ce{PH3} profile at high pressure, but the inclusion of the upper atmosphere, different mixing profiles and T-p profiles affects the \ce{P2} abundance in the HD 189733b case differently to the \citet{Wang2017} T$_{\rm eq}$ = 1000 K profile.

The impact of photochemistry on PHO chemistry is stark from Figure \ref{fig:HD189}. 
For both metallicity cases, the larger molecules are photodisocciated, leaving \ce{P2} and PO as the P carrying species in the middle and upper atmosphere. 
This is because photochemistry directly breaks down or produces radicals (primarily H) that leaves behind only simple, small molecules with relatively strong bonds.
\ce{PH3} is severely depleted from the upper atmosphere through photochemical effects, with PH now being the most abundant hydrogen bearing P molecule, suggesting that photochemistry induces a cascade from \ce{PH3} to \ce{PH2} and PH, also commonly seen for \ce{NH3} and \ce{CH4} photochemistry.
For the 10 times Solar case, the initial abundances of HOPO and \ce{P2} are reduced by photochemistry leaving a PO dominated atmosphere, this also suggests radical formation, such as H which is in high abundance in the mid-upper atmosphere, that destroys HOPO (Section \ref{sec:disc}).
We discuss the key chemical pathways that give rise to the results in Section \ref{sec:disc} and potential observational aspects of these results in Section \ref{sec:obs_cons}.

\section{Application to warm Neptune atmospheres}
\label{sec:wn}

\begin{figure*}
    \centering
    \includegraphics[width=0.49\textwidth]{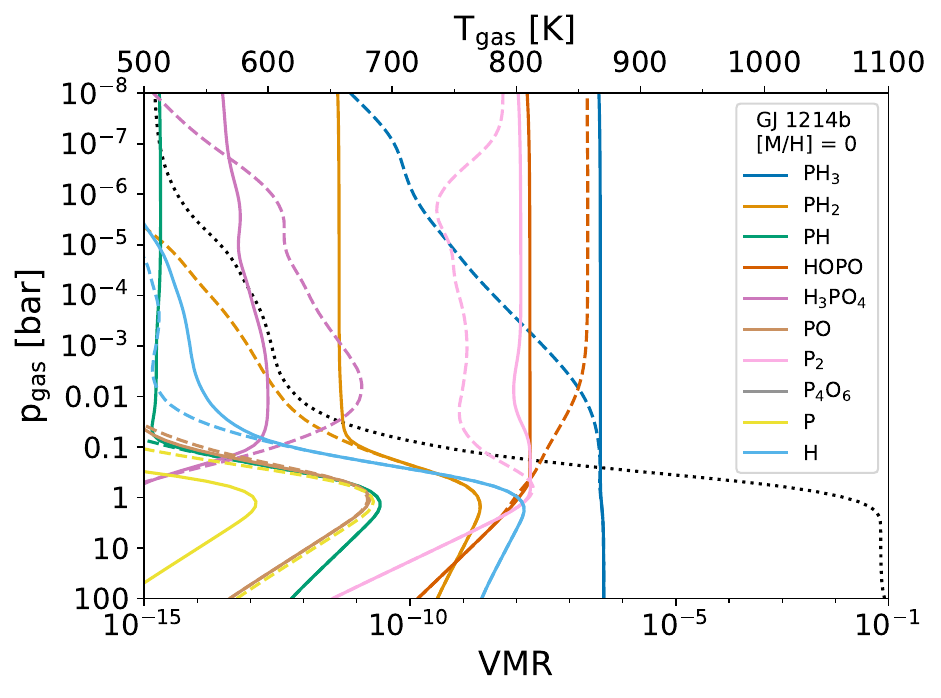}
    \includegraphics[width=0.49\textwidth]{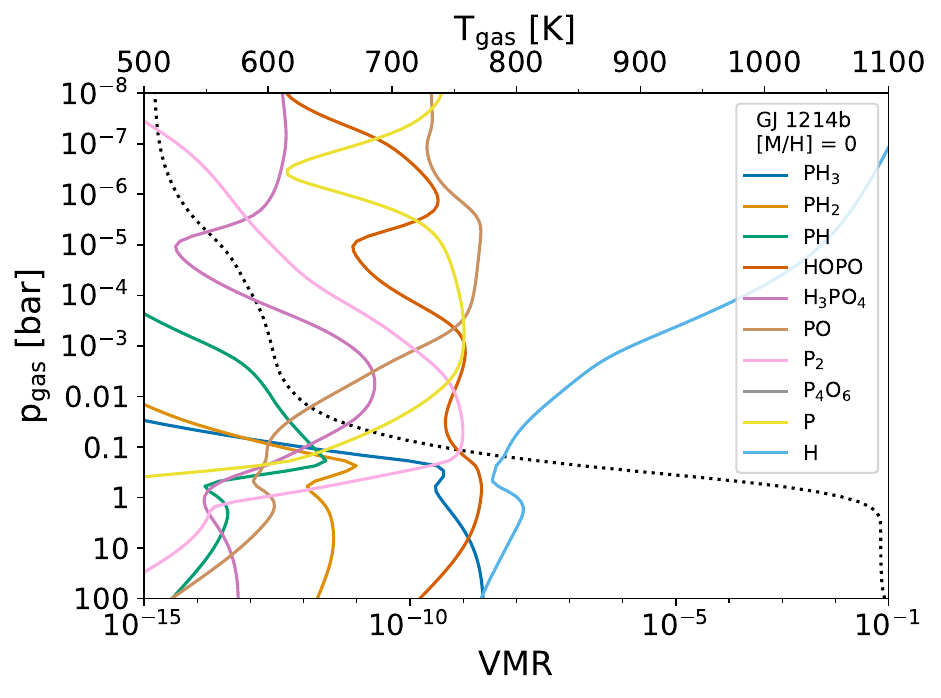}
    \includegraphics[width=0.49\textwidth]{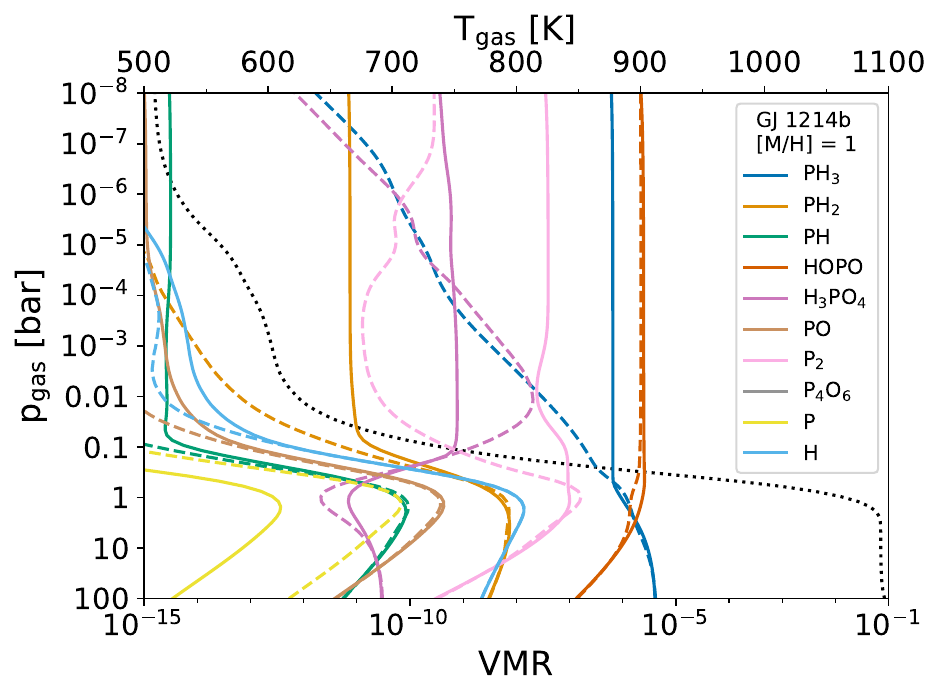}
    \includegraphics[width=0.49\textwidth]{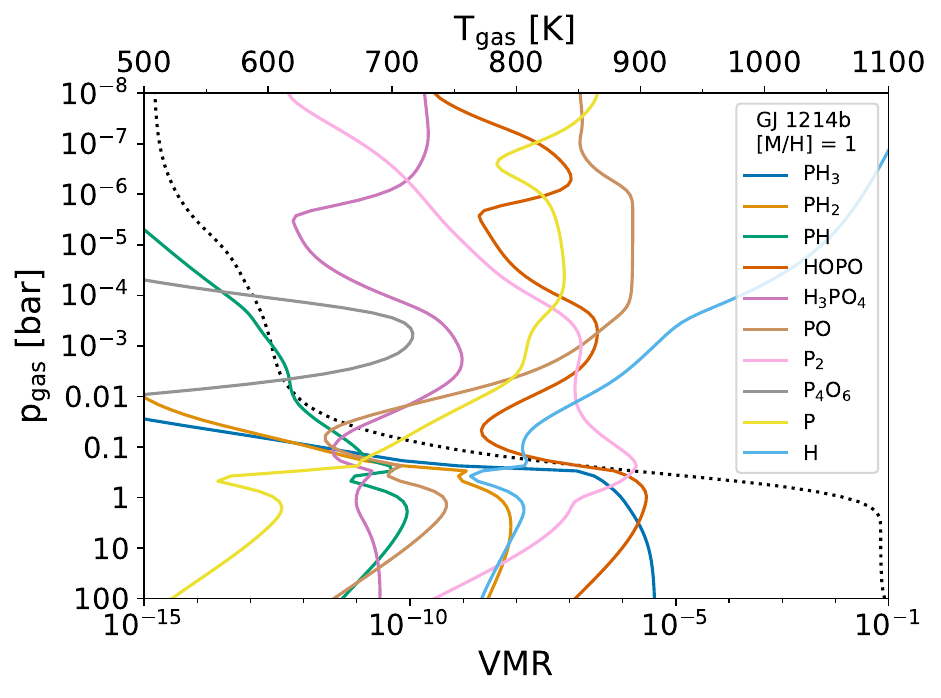}
    \includegraphics[width=0.49\textwidth]{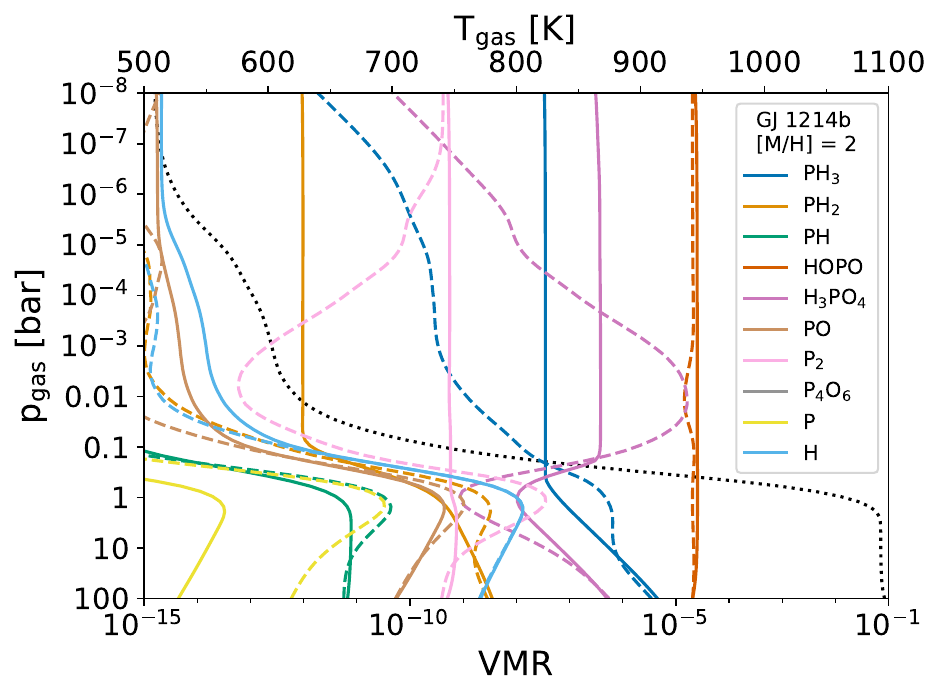}
    \includegraphics[width=0.49\textwidth]{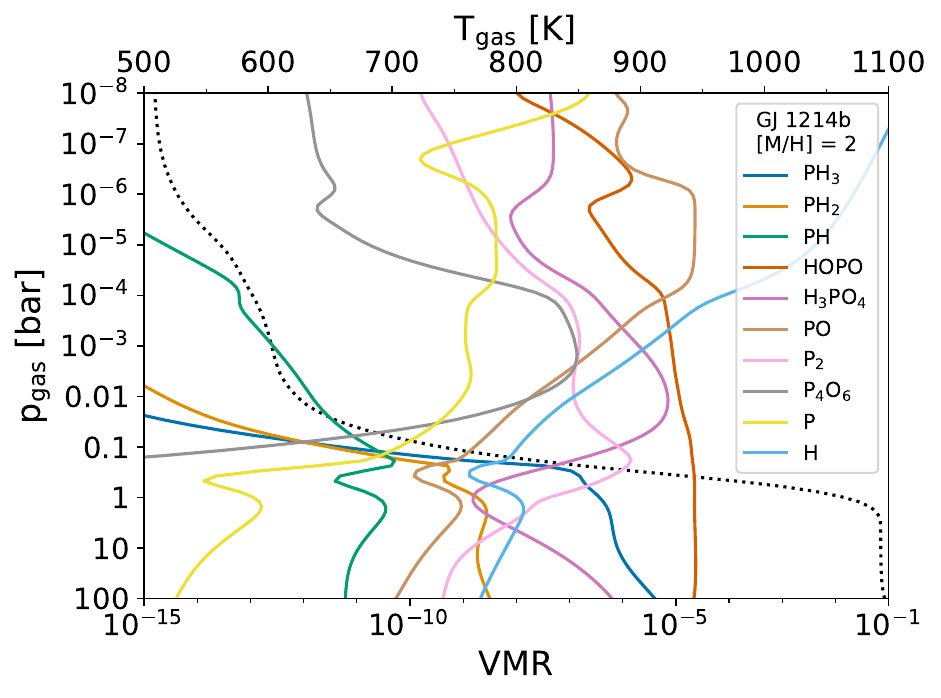}
    \caption{VULCAN PHO network results for the benchmark GJ 1214b 1D case, with the \citet{Moses2022} K$_{\rm zz}$ expression.
    The volume mixing ratio (VMR) of each species is shown as solid lines and the T-p profile as a black dotted line.
    The dashed lines show the chemical equilibrium values for each species.
    Top left: thermochemistry only at [M/H] = 0. 
    Top right: photochemical model at [M/H] = 0.
    Middle left: thermochemistry only at [M/H] = 1. 
    Middle right: photochemical model at [M/H] = 1.
    Bottom left: thermochemistry only at [M/H] = 2.
    Bottom right: photochemical model at [M/H] = 2.
     }
    \label{fig:GJ1214b}
\end{figure*}

In this section, we apply the PHO network to the GJ 1214b system as a representative warm Neptune atmosphere.
We calculate a global average T-p profile for GJ 1214b using the HELIOS radiative-convective equilibrium (RCE) model \citep{Malik2017}, which is then used as input to the VULCAN model.
We examine Solar, 10 times Solar and 100 times Solar metallicity cases and follow the K$_{\rm zz}$ profile expression from the \citet{Moses2022} (their Eq. 1) study, scaled to the properties of GJ 1214b (H$_{\rm 1 mbar}$ = 209 km, T$_{\rm eff}$ = 679 K).
This leads to a K$_{\rm zz}$ profile similar to \citet{Moses2022}'s Figure 2. T$_{\rm eff}$ = 700 K.
Figure \ref{fig:T_p_Kzz} shows the input T-p and K$_{\rm zz}$ profile.

Figure \ref{fig:GJ1214b} presents the GJ 1214b test cases.
For the thermochemical kinetics only cases without photochemistry, \ce{PH3} is dominant only in the Solar-metallicity case, while being replaced by HOPO in the higher-metallicity cases.
Species are quenched around the 0.1 bar pressure level in all cases, leading to strong non-equilibrium behaviour at pressures probed by transmission and emission. 
The higher abundance of HOPO at chemical equilibrium at higher metallicites along with the quenching behaviour, contributes to its ubiquity in the upper atmosphere.
For the photochemical cases, as in the HD 189733b case, the effects are striking, again, HOPO, PO and \ce{P2} tend to dominate most of the upper atmosphere, with \ce{PH3} being confined to its chemical equilibrium abundances in the deep atmosphere.
The production of H radicals in the upper atmosphere due to photochemical processes, promotes the destruction of the initial HOPO, leading to a PO dominated composition.
This large H radical production is not present in the non-photochemical models.
We discuss the key chemical pathways that give rise to the results in Section \ref{sec:disc} and potential observational aspects of these results in Section \ref{sec:obs_cons}.

\section{Application to deep Jupiter and Saturn atmospheres}
\label{sec:djs}

\begin{figure*}
    \centering
    \includegraphics[width=0.49\textwidth]{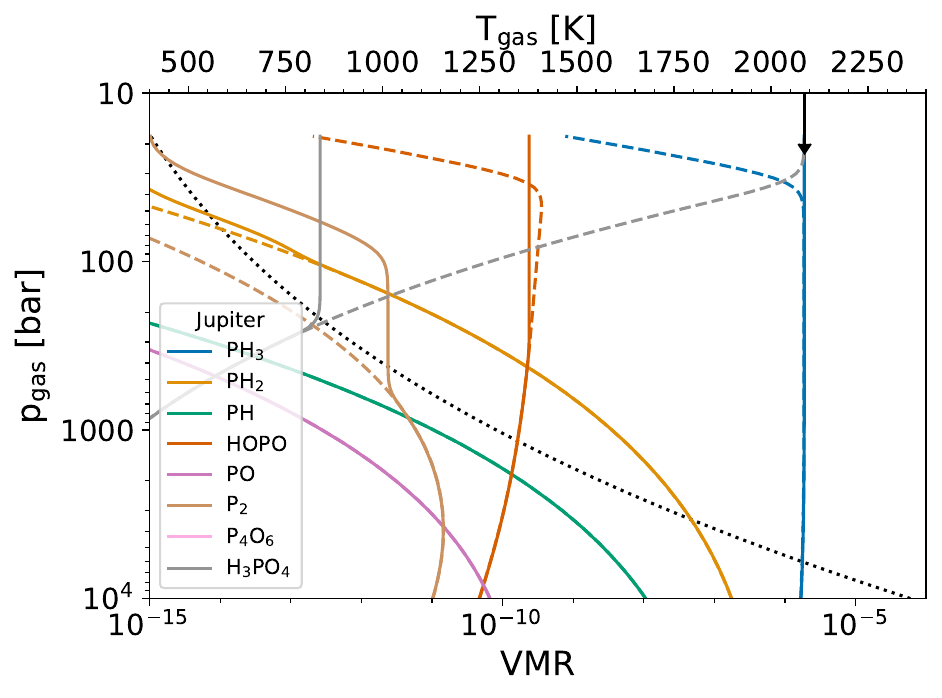}
    \includegraphics[width=0.49\textwidth]{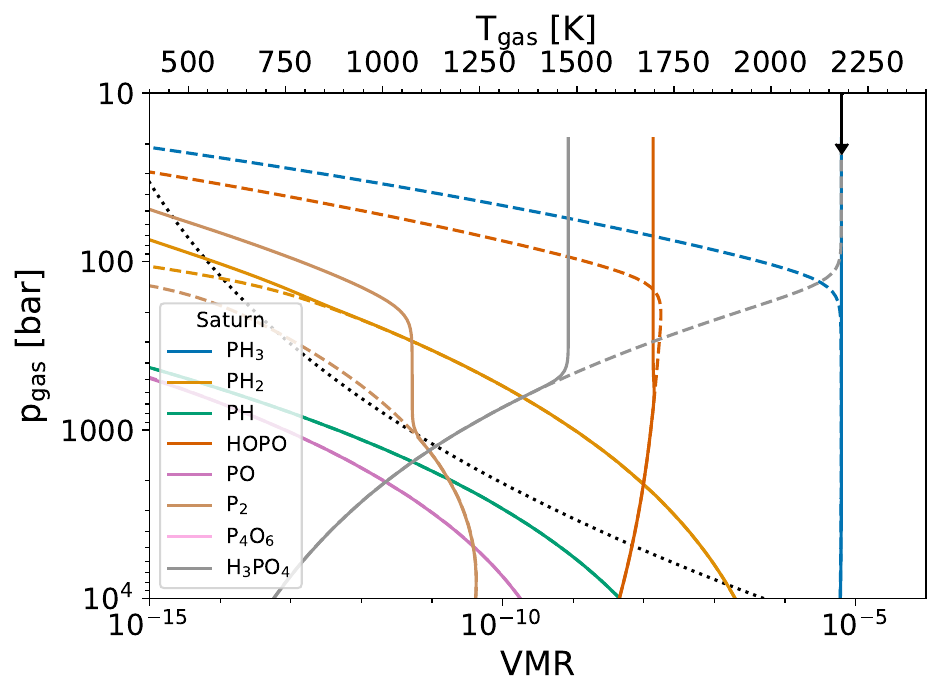}
    \caption{VULCAN PHO network results for the deep Jupiter (left) and Saturn (right) with the T-p profile from \citet{Moses2005} extended to 10$^{4}$ bar assuming an adiabat, and assuming a constant K$_{\rm zz}$ = 10$^{8}$ cm$^{2}$ s$^{-1}$ \citep{Wang2016}.
    The volume mixing ratio (VMR) of each species is shown as solid lines and the T-p profile as a black dotted line. 
    The dashed lines denote the chemical equilibrium abundances.
    The black arrow denotes the \ce{PH3} abundance retrieved by \citet{Fletcher2009} at lower pressure regions.
     }
    \label{fig:djs}
\end{figure*}

In this section, the network is applied to the deep atmospheres of Jupiter and Saturn.
We take the T-p profiles for both gas giants from \citet{Moses2005}, following an adiabatic profile to extend it to 10$^{4}$ bar, and assume a constant K$_{\rm zz}$ = 10$^{8}$ cm$^{2}$ s$^{-1}$, following \citep{Wang2016}.
For Jupiter, we take the P and O abundances from Table 1 in \citet{Mousis2021}, specifically the O ratio (1450 ppm) from \citet{Li2020} and P ratio (1.08 ppm) from \citet{Fletcher2009}. 
For Saturn, we take the P ratio value from \citet{Atreya2020} (3.64 ppm) and the O ratio from \citet{Cavalie2024}, which was estimated to be around eight times the solar values of \citet{Lodders2021} (4100 ppm).
We take He ratios for both planets from the \citet{Atreya2020} review.

In Figure \ref{fig:djs}, we present the results for the Jupiter and Saturn profiles, which show an interesting dynamic: the initial chemical equilibrium abundance of \ce{H3PO4} decreases from its initial value, because of dissociation into \ce{HOPO2}.
Eventually, \ce{H3PO4} becomes a negligible species in both atmospheres. 
\ce{H3PO4} is quenched at a pressure level of around 300 bar at these low abundances, which allows \ce{PH3} to form and mix upward to the upper atmosphere from its initial CE abundance to its observed abundance \citep{Fletcher2009}.
\ce{P4O6} is produced in negligible amounts in both models.

\section{Discussion} 
\label{sec:disc}

In this section, we discuss aspects of the PHO network, and the main chemical mechanisms that drive our results.
In addition, we discuss shortcomings of the model and gaps in the network that can be expanded and addressed with further experimental and theoretical efforts.

\subsection{Oxygenation mechanisms}

In this section, we describe the formation mechanisms of the small oxygenated molecules that are most ubiquitous in the simulations.
We focus on the formation of \ce{P2}, PO and HOPO as they are the main products of the network.

\subsubsection{Small oxygenated phosphorus species}

\begin{figure*}
    \centering
    \includegraphics[width=0.75\linewidth]{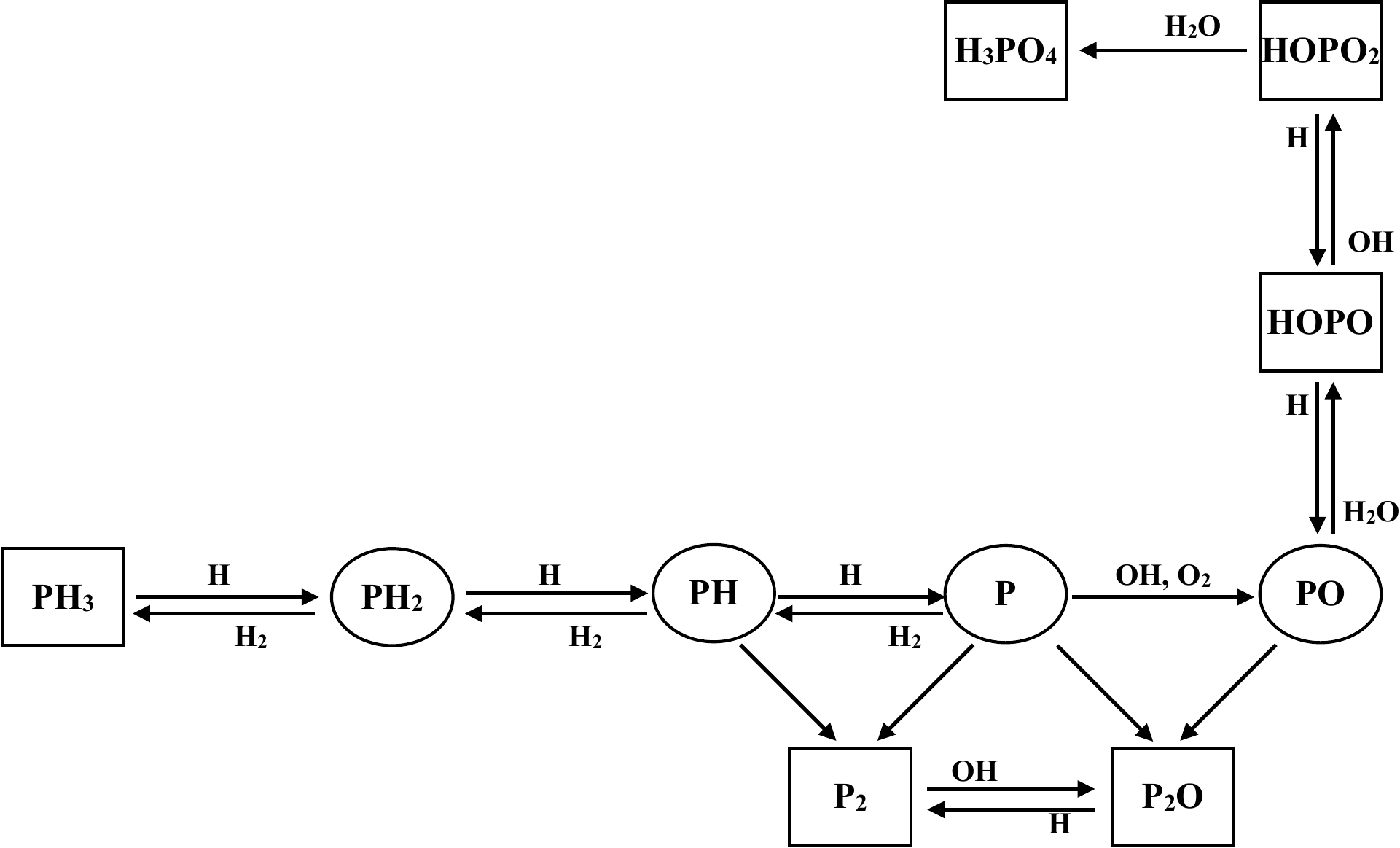}
    \caption{Diagram showing the main chemical pathways present in the PHO scheme between key species.}
    \label{fig:P_diag}
\end{figure*}

A key result from our simulations is that small oxygenated phosphorus species play a major role in the phosphorus chemistry in exoplanet atmospheres, in particular HOPO and PO.
The only exoplanet simulation where \ce{PH3} is dominant in the upper atmosphere is in the GJ 1214b case without photochemistry. 
Our results suggest that metal enhancement and photochemistry efficiently produce HOPO, PO and \ce{P2}, which are found in larger abundance than \ce{PH3}.

In the network, the formation of \ce{P2} follows a simple path from \ce{PH3} and subsequent reactions with radicals e.g.
\begin{align*} 
    2(\ce{PH3} +\ce{H} &\rightarrow \ce{PH2} + \ce{H2}) \\
    2(\ce{PH2} + \ce{H} &\rightarrow \ce{PH} + \ce{H2}) \\
    \ce{PH} + \ce{H} &\rightarrow \ce{P} + \ce{H2} \\
    \ce{P} + \ce{PH} &\rightarrow \ce{P2} + \ce{H}\\
    2\ce{H2} &[M]\rightarrow 4\ce{H} \\
    net: 2\ce{PH3} &\rightarrow \ce{P2} + 3\ce{H2} \numberthis \label{eq:P2_th}
\end{align*} 

This pathway enables a rapid formation of \ce{P2} in the atmosphere, especially when photochemistry contributes to the dissociation of hydrogen and produces H radicals which can further dissociate PH$_{x}$ molecules through mechanism \ref{eq:P2_th}.
\ce{P2} is also naturally favoured at chemical equilibrium in the hot Jupiter HD 189733b models, as shown by the thermochemical model results.
In cooler atmospheres that experience low irradiation, such as Jupiter, \ce{PH2} + \ce{PH2} + M = \ce{P2H4} + M can compete with other loss processes for \ce{PH2}, and if \ce{P2H4} condenses, the phosphorus will be locked into the condensed state.

Formation of PO follows a similar route but with odd-oxygen radicals (OH or O) reacting with P e.g.
\begin{align*} 
    \ce{PH3} + \ce{H} &\rightarrow \ce{PH2} + \ce{H2} \\
    \ce{PH2} + \ce{H} &\rightarrow \ce{PH} + \ce{H2} \\
    \ce{PH} + \ce{H} &\rightarrow \ce{P} + \ce{H2} \\
    \ce{P} + \ce{OH} &\rightarrow \ce{PO} + \ce{H} \\ 
    \ce{H2O} + \ce{H} &\rightarrow \ce{OH} + \ce{H2} \\
    net: \ce{PH3} + \ce{H2O} + 3\ce{H} &\rightarrow \ce{PO} + 4\ce{H2}\numberthis \label{eq:PO_ph}
\end{align*} 
The last oxygenation steps are very rapid reactions \citep{Douglas2022}, allowing efficient formation of PO when odd-oxygen radicals are available in the atmosphere.
This is particularly true when photochemistry is occurring, explaining the large amount of PO produced in the photochemical simulations of both hot Jupiters and warm Neptunes.

HOPO forms directly from PO as noted in \citet{Wang2016}, and with the derived rate from \citet{Jayaweera2005}
\begin{align*}
    \ce{PO} + \ce{H2O} &\rightarrow \ce{HOPO} + \ce{H} \numberthis \label{eq:HOPO_ph}
\end{align*} 
The consequences of this reaction are seen primarily in the photochemical models, where in the upper atmosphere HOPO is broken down by radicals to produce PO, while in the deeper atmosphere HOPO is retained.
There is generally a transition region between PO and HOPO in the middle atmosphere, where fewer radicals are being produced compared to the upper atmosphere, and some HOPO survives.
As PO diffuses downward, it reacts with \ce{H2O}, forming HOPO, while as HOPO diffuses upwards, it is broken down into PO by radicals.
Without radicals to break down HOPO, HOPO remains the primary P bearing species as seen in our enhanced metallicity thermochemical kinetics only simulations.
PO remains a minor species in these thermochemical kinetics only simulations.

In Figure \ref{fig:P_diag} we present a diagram of the main chemical pathways between key species in the network.
This shows the main routes to produce the small oxygenated molecules starting from \ce{PH3}.
These follow similar pathways to those in \citet{Wang2016}.

Overall, the network characteristics show a simple and direct path of conversion of \ce{PH3} to \ce{P2}, HOPO and PO for the Solar metallicity models, especially when photochemical processing occurs.
At higher metallicities, the balance between HOPO and PO is key to understanding the chemical profile, with photochemical processing playing a major role in determining the transition zones between the two molecules.
Without photochemistry, HOPO remains the dominant molecule in these atmospheres.

\subsubsection{Formation of \ce{H3PO4}}

Formation of the end product \ce{H3PO4} is driven by the reaction from \citet{Douglas2022} (\ce{HOPO2} + \ce{H2O} + M $\rightarrow$ \ce{H3PO4} + M), making the abundance of \ce{H3PO4} highly dependent on the local \ce{HOPO2} and \ce{H2O} availability. 
In all thermochemical models, \ce{H3PO4} remains a minor species except for the cool, metal enriched systems such as the GJ 1214b 100 times Solar model.
This suggests \ce{H3PO4} is only present in metal enriched scenarios, where the oxygenation process can provide the HOPO and then \ce{HOPO2} needed to produce \ce{H3PO4}
This is further evidenced by the GJ 1214b 100 times Solar photochemical model, where \ce{H3PO4} occurs at moderate abundance in the mid atmosphere.
This suggests an efficient formation pathway to the end product \ce{H3PO4} when local thermochemical conditions are suitable.
The formation of \ce{HOPO2} is primarily driven by reaction of oxygen radicals with HOPO and other oxygenated phosphorus oxides such as PO and \ce{PO2}; these species are produced in generally higher quantities through photochemical processing, especially at high metallicity.

We caution that we have not included any other pathways for the formation and destruction of \ce{H3PO4}, and as noted in Sect. \ref{sec:djs}, we lack high pressure rate data for the reaction involving \ce{H3PO4}, which could level off the rate of formation of this molecule at moderate to high pressures.

\subsubsection{Formation of \ce{P4O6}}

Our results suggest the formation of \ce{P4O6} end product is highly unfavourable in all thermochemical environments.
It occurs maximally at the parts per billion level in the highly metal enriched and photochemical environments.
However, we again caution that we have only explored a single \ce{P4O6} formation pathway.

\subsection{Data gaps and needs}

In this section, we discuss the current gaps in the PHO network and potential areas of improvements.
Several general areas of uncertainty remain for the kinetics of P chemistry are listed below.
\begin{itemize}
    \item Many rate coefficients are theoretical estimates. While simple recombination reactions can probably be calculated reasonably accurately (within a factor of 2), reactions over complex potential energy surfaces involving barriers are much more uncertain.
    \item Sources and derivations of rate data are not fully consistent across the species list.
\end{itemize}

More specifically, we highlight below reactions with significant sensitivity in the model, where improved estimates of rate coefficients would be particularly beneficial, as well as additional reaction pathways.
\begin{itemize}
    \item Reactions that build \ce{P4} (e.g. \ce{P2} + \ce{P2} + M).
    \item Reactions that build \ce{P2H2} and \ce{P2H4} (e.g. PH + PH + M and \ce{PH2} + \ce{PH2} + M), which are both molecules of atmospheric interest, especially in cold reducing environments.
    \item Bimolecular reactions involving radicals interacting with phosphorus oxides such as PO, \ce{PO2} and \ce{PO3}.
    \item Additional pathways for building larger oxidised molecules such as \ce{H3PO4} and \ce{P3O4}.
    \item Several high pressure rates are unknown for important combination reactions such as OH + PO + M $\rightarrow$ HOPO + M.
    \item We lack the high pressure rate for the reaction that forms \ce{H3PO4}.
\end{itemize}

On the photochemistry side, several aspects of data are missing or incomplete.
\begin{itemize}
\item Photodissosation of larger molecules such as \ce{H3PO4} is not included.
\item Only one photolysis product branch is given for each phosphorus molecule with unknown quantum efficiencies.
\end{itemize}

Overall, the PHO network remains highly approximate, with many reactions containing uncertain and estimated rates.
Significant effort will be required to experimentally and theoretically build a more reliable and sound PHO network, however, our current study provides a useful guide into what mechanisms require the most attention going forward.

\subsection{Combining with the SNCOH network}

In this study, we have focused on the PHO system exclusively, ignoring the impact of S, N and C species on the P chemistry, which may be significant.
However, our PHO only effort allows us to analyse the main properties and mechanisms of the proposed PHO photochemical scheme without interference from other species.
Several additions will have to be made to properly integrate the scheme into the SNCOH network.
Of note, two important molecules and their pathways in the full SNCOHP network to include are PS and PN, where
some reaction rates of PN species are available in \citet{Douglas2022}.
These aspects will be explored in a follow up paper.

We can expect several effects on the P species from adding S, N and C species.
For example, the addition of SNCOH will affect the impact of photochemistry on P species through increased UV shielding. This would reduce the effectiveness and depth that photolysis of P products occurs, possibly changing the vertical profiles of P species.
In particular, the boundary between HOPO and PO may change due this affect, as changes in the H radical vertical profile occur with the addition of other species.

Reactions with S, N and C radicals with P species will produce PN, PS and CP complexes, possibly reducing the amount of HOPO and PO seen in the simulations with the PHO only network.
In addition, more H radicals may be present at deeper depths when additional molecules such as \ce{NH3} and \ce{H2S} are included, greatly affecting the chemical structure of the atmosphere. 
This may reduce the HOPO and PO to below ppm levels, making it harder to detect in these atmospheres.
Overall though, we expect similar P species (HOPO, PO and \ce{P2}) to be produced with the full network, and our main conclusions regarding the chemical mechanisms should not be significantly affected.

\section{Observational consequences}
\label{sec:obs_cons}

\begin{figure}
    \centering
    \includegraphics[width=0.49\textwidth]{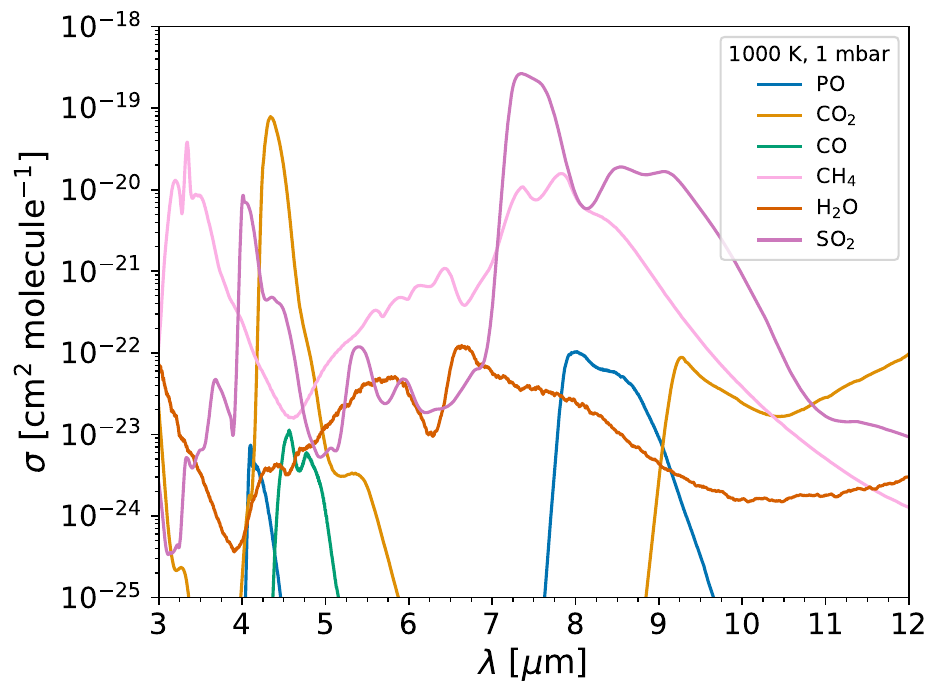}
    \caption{Cross sections of PO, CO$_{2}$, CO, CH$_{4}$, H$_{2}$O and SO$_{2}$ at 1000 K and 1 mbar.
    Features are seen at around 4.1 $\mu$m and 8$\mu$m, which lie in the wavelength range of NIRSpec, NIRCam and MIRI JWST instruments, commonly used for exoplanet atmosphere characterisation.
    However, these may be obscured by CH$_{4}$ and SO$_{2}$.}
    \label{fig:PO}
\end{figure}

Our results suggest that for hot Jupiters like HD 189733b, \ce{PH3} will not be detectable in both transmission and emission with current instrumentation, because \ce{PH3} remains below the ppm level at pressure levels probed by transmission and emission ($\sim$1-10$^{-4}$ bar).
Only in the deep atmosphere does the \ce{PH3} abundance rise above the ppm level.
We find photochemical processing creates a large PO abundance, especially at enhanced metallicities.
This makes PO a promising molecule to detect with JWST and other telescopes.

For warm Neptunes like GJ 1214b, a similar picture emerges, where HOPO, and PO remain the strongest P bearing species to detect, especially at higher metallicities and with photochemical processing.
\ce{P2} is a homonuclear molecule, and is probably does not have strong absorption features.
Atomic P appears in the very upper atmosphere in some cases, but this only has strong lines at UV wavelengths \citep[e.g.][]{Kurucz1995}.
Our results suggest \ce{PH3} will also not be detectable in transmission or emission for this class of planet.

Overall, our study suggests that the P bearing species of interest for exoplanet characterisation are HOPO and PO, of which only PO currently has line-list data \citep{Prajapat2017}.
In Figure \ref{fig:PO}, we present cross-sections of PO produced using the \citet{Prajapat2017} line-list data at 1000 K and 1 mbar, compared to other molecules of interest found in hot Jupiter and warm Neptune exoplanets.
These show features in bands at around 4.1$\mu$m and 8$\mu$m, which are probed by the commonly used NIRSpec G395H, NIRCam Grisim and MIRI LRS JWST modes, suggesting that observational evidence for PO may already be present in current JWST data for metal enhanced planets.
However, the 4.1$\mu$m PO band would be obscured by the presence of SO$_{2}$, CH$_{4}$ as well as CO$_{2}$, which have much larger cross sections in that wavelength range.
PO may fill in the gap between the SO$_{2}$ feature and the ramp in opacity of the CO$_{2}$ feature, leading to an apparent steeper climb in opacity near 4.1$\mu$m compared to just SO$_{2}$ and CO$_{2}$ alone.
A promising distinguishing feature is the 8$\mu$m band for PO, which would be clearly apparent above the H$_{2}$O opacity and fill in the gap between the SO$_{2}$ bands and appear to broaden the 7.5$\mu$m SO$_{2}$ feature.
If CH$_{4}$ is present, it is likely to dwarf any PO signal in these JWST wavelength ranges.

The formation of SO$_{2}$ is favoured at metallicities around 10 times Solar \citep{Tsai2023}, which is also the range where the PHO network produces PO at ppm levels. 
This suggests SO$_{2}$ and PO may form together as photochemical products in this range for hot Jupiters and warm Neptunes.

\section{Summary and Conclusions} 
\label{sec:conc}

In this study, we present a PHO photochemical network for exoplanet atmospheres by updating the \citet{Wang2016} PHO network with new reaction rates sourced from the literature and with new theoretical calculations.
We also add ten photochemical reactions that impact P bearing species, some with new theoretical UV cross-sections.
Overall, we were able to update around 25$\%$ of the \citet{Wang2016} network, improving the robustness of the overall network substantially.
For the first time, we explored a \ce{P4O6} formation mechanism through calculating theoretical rates for the termolecular recombination reaction (\ce{P2O3} + \ce{P2O3} + M $\rightarrow$ \ce{P4O6} + M), but find the \ce{P4O6} abundance to be a negligible component in the atmospheres simulated.

Overall, our results suggest, for hot Jupiters and warm Neptunes, HOPO, PO, \ce{P2} and atomic P are the key P bearing species, especially at higher metallicities and where photochemical processing is present.
Our results suggest \ce{PH3} is only seen in Solar metallicity, cold planets where photochemistry is negligible, as well as cold planets with similar O and P ratios to Jupiter and Saturn.
We suggest that retrieval models include PO as part of their species detection suite and include HOPO when line-lists or opacity data become available.
Due to the spectral features of PO, this molecule may already be traceable in current JWST NIRSpec, NIRCam and MIRI transmission and emission spectra data of metal enhanced planets.

Despite our progress, we caution that our proposed PHO network contains many approximate rate coefficients and potentially missing key reaction pathways, and so strong conclusions regarding the abundance predictions from these simulations should be considered carefully. 
These concerns will need to be addressed through future experiment and theoretical calculations to put phosphorus kinetics on a firmer footing.

Our study points to the importance of considering photochemistry for P networks and provides physical mechanisms for consideration when interpreting observational data for \ce{PH3} (non-)detection.
Due to \ce{PH3}'s status as a biomarker molecule, improving the accuracy of phosphorus kinetic networks through experimental and/or theoretical efforts will be an important goal for the exoplanet field going into the near future.
Our PHO study forms the basis for our future combined PSCHNO photochemical network.

\begin{acknowledgments}
E.K.H. Lee is supported by the SNSF Ambizione Fellowship grant (\#193448).
S-M. Tsai is supported by the University of California at Riverside.
J.M.C. Plane was supported by the UK Science and Technology Facilities Council (grant ST/P000517/1).
J. Moses and C. Visscher were supported by NASA Exoplanet Research Program grant 80NSSC22K0314.
This material is based in part on work supported by the U.S. Department of Energy, Office of Science, Office of Basic Energy Sciences, Division of Chemical Sciences, Geosciences, and Biosciences under contract No. DE-AC02-06CH11357.
We thank P. Rimmer for advice on the ARGO Venus phosphorus photochemical model.
\end{acknowledgments}

\vspace{5mm}

\software{VULCAN \citep{Tsai2017,Tsai2021}}

\bibliography{bib}{}
\bibliographystyle{aasjournal}

\appendix

\section{Theoretical contributions to the PHO network}
\label{app:P4O6}

\subsection{Theoretical Kinetics Calculation for PxHy Reactions}

Ab initio transition state theory (TST) was used to predict the rate constants for the abstractions $^{3}$PH + H $\rightarrow$ $^{4}$P + \ce{H2}; $^{2}$\ce{PH2} + H $\rightarrow$ $^{3}$PH + \ce{H2}; \ce{PH3} + H $\rightarrow$ $^{2}$\ce{PH2} + \ce{H2}; and $^{3}$PH + $^{2}$\ce{PH2} $\rightarrow$ $^{4}$P + \ce{PH3}. 
The rovibrational properties of the stationary points on these potential energy surfaces were evaluated at the CCSD(T)/cc-pV(Q+D)Z level. 
The barrier heights were evaluated with a composite approach that combined (i) a CCSD(T) complete basis set (CBS) limit obtained from extrapolation of cc-pV(5+D)Z and cc-pV(6+D)Z energies, (ii) CCSDT(Q)/cc-pV(D+D)Z corrections for higher order excitations, (iii) and CCSD(T)/CBS core-valence corrections from all electron calculations for TZ and QZ basis sets. 
The partition functions were evaluated within the rigid-rotor harmonic-oscillator approximation. Asymmetric Eckart tunneling corrections were also included.

The radical-radical recombination of \ce{PH2} with \ce{PH2} was treated with variable reaction coordinate (VRC)-TST.
A direct sampling CASPT2/cc-pV(T+D)Z approach was used to evaluate the interaction energies in the transition state region. 
One-dimensional P-P distance dependent corrections were obtained from the combination of a geometry relaxation correction and a complete basis set limit correction. 
The geometry relaxation correction was obtained from constrained geometry evaluations at the CASPT2/cc-pV(Q+D)Z level. 
The basis set relaxation correction was obtained from extrapolation of CASPT2/cc-pV(5+D)Z and CASPT2/cc-pV(6+D)Z evaluations along the CASPT2/cc-pV(Q+D)Z minimum energy path. 
A dynamical correction of 0.85 was applied to the final VRC-TST predictions.

Pressure dependent predictions for the $^{2}$\ce{PH2} + $^{2}$\ce{PH2}  $\rightarrow$ \ce{P2H4} $\rightarrow$  \ce{PPH2} + \ce{H2} system were obtained from one-dimensional master equation simulations incorporating the VRC-TST flux for the recombination channel. 
The remaining channels were treated as described above for the abstraction reactions. One-dimensional hindered rotors were included as appropriate. 
The energy transfer rates were treated within the exponential down formalism and Lennard-Jones collision rates.

The PH + H and \ce{PH2} + H recombination reactions were similarly treated with VRC-TST, but now employing multi-reference configuration interaction MRCI+Q based evaluations for the direct sampling over the  interaction potential. 
These direct evaluations included the Davidson correction for higher order interactions and were performed for the aug-cc-pV(T+D)Z basis. 
A dynamical correction of 0.9 was applied to the final VRC-TST predictions. 
One-dimensional master equation simulations were again used to predict the pressure dependence, with the binding energies determined from equivalent CCSD(T) based composite methods. 

\subsection{Theoretical Calculations of Cross Sections and PxOyHz Rate Constants}

\begin{figure}
    \centering
    \includegraphics[width=0.5\textwidth]{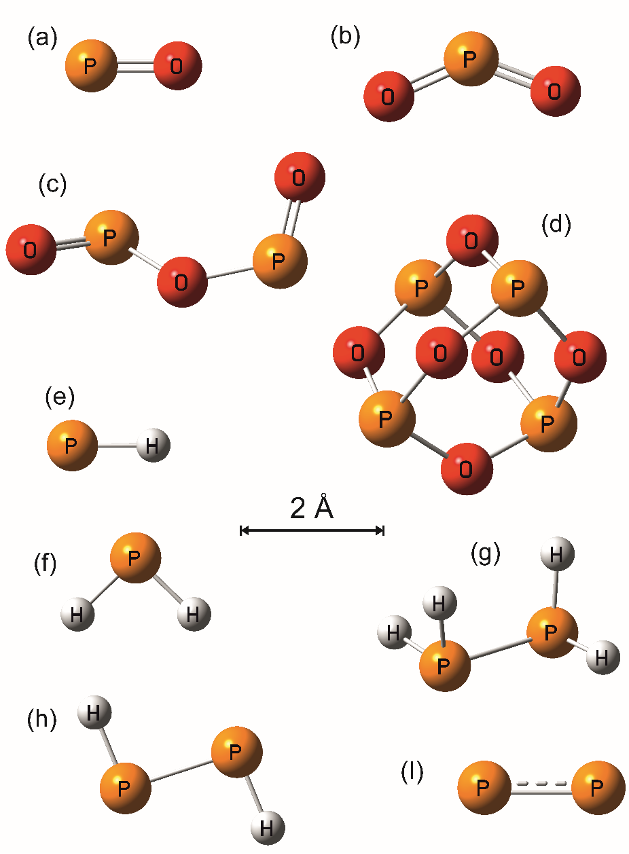}
    \caption{Molecular geometries of (a) PO, (b) \ce{PO2}, (c) \ce{P2O3}, (d) \ce{P4O6}, (e) PH, (f) \ce{PH2}, (g) \ce{P2H4}, (h) \ce{P2H2} and (i) \ce{P2}.}
    \label{fig:PHO_geo}
\end{figure}

\begin{table*}[ht]
    \label{tab:QM}
    \centering
    \caption{Molecular properties of the stationary points on the potential energy surfaces for PO + OPO and \ce{P2O3} + \ce{P2O3} and some relevant P$_{\rm x}$H$_{\rm y}$ species. $^{a}$ Calculated at the B3LYP/6-311+g(2dp) level of theory. $^{b}$ Enthalpy and Gibbs free energy of formation at 298 K calculated at the G4 level of theory and using reference values for P(g) and H(g) of $\Delta_fH^\circ$ (298 K) = 316.39 and 218.00 kJ mol$^{-1}$ respectively \citep{Chase1998}.}
    \begin{tabularx}{\textwidth}{l|b|X|X|s|s}
        \toprule
        Molecule & Geometry  \newline (Cartesian \newline co-ordinates in \AA)$^{a}$ & Rotational \newline constants \newline (GHz)$^{a}$ & Vibrational \newline frequencies  \newline (cm$^{-1}$)$^{a}$ & $\Delta_fH^\circ$ \newline (298 K)  \newline (kJ mol$^{-1}$)$^{b}$ & $\Delta_fG^\circ$ \newline (298 K)  \newline (kJ mol$^{-1}$)$^{b}$ \\
        \midrule
        PO & P 0.0 0.0 0.0585 \newline O 0.0 0.0 1.5415 & 21.78804 & 1240 & -41.14 & -63.36 \\
        \midrule
        \ce{PO2} & P 0.087 0.0 0.041 \newline O 0.032 0.0 1.5156 \newline O 1.182 0.0 -0.949 & 97.5896 \newline 8.54002 \newline 7.85283 & 382 \newline 1059 \newline 1304 & -286.2 & -288.9 \\
        \midrule
        \ce{P2O3} & P 1.301 -0.306 -0.456 \newline P -1.633 0.378 0.039 \newline O 2.472 0.337 0.147 \newline O -0.024 0.692 -0.168 \newline O -1.843 -1.044 0.347 & 12.5877 \newline 1.70504 \newline 1.56174 & 34 \newline 101 \newline 132 \newline 329 \newline 442 \newline 609 \newline 784 \newline 1256 \newline 1279 & -644.5 & -630.4 \\
        \midrule
        \ce{P4O6} & O -1.537 0.935 0.209 \newline O -2.835 -1.257 0.219 \newline O -0.291 -1.284 0.070 \newline O -1.451 -2.068 2.200 \newline O -0.151 0.124 2.190 \newline O -2.696 0.150 2.339 \newline P 0.014 -1.457 1.697 \newline P -2.951 -1.427 1.872 \newline P -1.599 -0.512 -0.611 \newline P -1.437 1.129 1.860 & 1.04977 \newline 1.04977 \newline 1.04977 & 267 (×3) \newline 294 (×2) \newline 388 (×3) \newline 533 (×3) \newline 564 (×3) \newline 588 \newline 621 (×3) \newline 640 (×2) \newline 712 \newline 908 (×3) & -1659 & -1531 \\
        \midrule
        PH & P 0.0 0.0 0.086 \newline H 0.0 0.0 1.514 & 254.08252 & 2347 & 238.2 & 211.4 \\
        \midrule
        \ce{PH2} & P 0.0 0.115 0.0 \newline H 1.021 -0.876 0.0 \newline H -1.0214 -0.876 0.0 & 271.85867 \newline 240.33917 \newline 127.56455 & 1130 \newline 2365 \newline 2373 & 138.6 & 124.5 \\
        \midrule
        \ce{P2H2} & P 0.450 1.229 1.470 \newline H 0.139 0.635 2.716 \newline P -0.207 -0.323 0.354 \newline H 0.153 0.269 -0.891 & 129.19206 \newline 7.53928 \newline 7.12357 & 613 \newline 694 \newline 780 \newline 979 \newline 2333 \newline 2349 & 124.6 & 113.7 \\
        \midrule
        \ce{P2H4} & P -0.361 0.249 -1.097 \newline H -0.277 -1.101 -1.518 \newline P -0.361 -0.249 1.097 \newline H 1.022 -0.491 1.287 \newline H 1.022 0.491 -1.287 \newline H -0.277 1.101 1.518 & 65.47798 \newline 5.67029 \newline 5.65539 & 198 \newline 418 \newline 630 \newline 653 \newline 815 \newline 885 \newline 1119 \newline 1126 \newline 2377 \newline 2387 \newline 2393 \newline 2401 & 37.98 & 60.18 \\
        \midrule
        \ce{P2} & P -2.086 0.682 0.0\newline P -0.195 0.812 0.0 & 9.088113 & 800 & 145.3 & 104.8 \\
        \bottomrule
    \end{tabularx}
\end{table*}

Electronic structure calculations were carried out using the Gaussian 16 suite of programs \citep{Frisch2016}. 
Vibrational frequencies, rotational constants and energies were calculated at the B3LYP/6-311+g(2d,p) level of theory, and energies using the very accurate G4 fourth-generation compound method of \citet{Curtiss2007}.
The Cartesian coordinates, molecular parameters, enthalpies and free energies of formation of the relevant phosphorus oxides and hydrides are listed in Table \ref{tab:QM}.
Their molecular geometries are illustrated in Figure \ref{fig:PHO_geo}. 

To calculate photodissociation spectra for PO, \ce{PO2}, \ce{P2}, \ce{P2H2} and \ce{P2H4}, their geometries were first optimized at the B3LYP/6-311+g(2d,p) level of theory \citep{Frisch2016}.
Vertical excitation energies and transition dipole moments for transitions from the ground state of each molecule to the first 30 electronically excited states were then calculated using time-dependent density function theory (TD-DFT) \citep{Bauernschmitt1996}.
\ce{P2H2} and \ce{P2H4} both photolyse at longer wavelengths by cleavage of the P--P bond, yielding PH + PH or \ce{PH2} + \ce{PH2} with photolysis thresholds of 338 nm and 508 nm, respectively (these thresholds correspond to the energy required to break the P––P bond). 
In the case of \ce{P2}, the photolysis threshold is 247 nm.
Rate coefficients for the recombination reactions PO + \ce{PO2} $\rightarrow$ \ce{P2O3} and \ce{P2O3} + \ce{P2O3} $\rightarrow$ \ce{P4O6} were calculated using the Master Equation Solver for Multi-Energy well Reactions (MESMER) program \citep{Glowacki2012}.
The internal energy of each species on the potential energy surface was divided into a contiguous set of grains (width 150 cm$^{-1}$) containing a bundle of rovibrational states, where the density of states was calculating using the relevant data in Table \ref{tab:QM}.
Each grain was then assigned a set of microcanonical rate coefficients for dissociation back to the reactants (PO + \ce{PO2}, or \ce{P2O3} + \ce{P2O3}) using an inverse Laplace transformation to link them directly to the high-pressure limiting recombination coefficients (k$_{\infty}$).
These coefficients were estimated using long-range transition rate theory \citep{Georgievskii2005} to be k$_{\infty}$(PO + \ce{PO2}) =  7.3 $\cdot$ 10$^{-10}$ (T/298)$^{0.167}$ cm$^{3}$ molecule$^{-1}$ s$^{-1}$ and k$_{\infty}$(\ce{P2O3} + \ce{P2O3}) =  8.4 $\cdot$ 10$^{-10}$ (T/298)$^{0.167}$ cm$^{3}$ molecule$^{-1}$ s$^{-1}$.
The exponential down model was used to estimate the probability of collisional transfer between grains.
The calculations were performed with \ce{N2} as the third body, where the average energy for downward transitions was set to $<\Delta$E$_{\rm down}>$ = 300 (T/298)$^{0.25}$ cm$^{-1}$ \citep{Gilbert1990}. 
The second-order recombination rates for the two reactions were calculated over large ranges of temperature (150 – 500 K) and pressure (10$^{-5}$–10$^{4}$ torr).
The low-pressure limiting rate coefficient for PO + \ce{PO2} is then k$_{0}$ = 4.56 $\cdot$ 10$^{-26}$ (T/298)$^{-4.25}$ cm$^{6}$ molecule$^{-2}$ s$^{-1}$, with a broadening factor F$_{\rm c}$ = 0.2. For \ce{P2O3} + \ce{P2O3}, k$_{0}$ = 2.37 $\cdot$ 10$^{-25}$ (T/298)$^{-2.99}$ cm$^{6}$ molecule$^{-2}$ s$^{-1}$, with a broadening factor F$_{\rm c}$ = 0.36.

\clearpage

\section{PHO reaction rate list}
\label{app:list}

In the following tables, we detail the included reactions inside the PHO network.
The photochemical reactions are detailed in Table \ref{tab:phchem}.
References for each reaction are indexed as follows: $^{a}$\citet{Jayaweera2005}, $^{b}$\citet{Douglas2019, Douglas2020, Douglas2022}, $^{c}$\citet{Twarowski1995,Wang2016}, $^{d}$\citet{Haworth2002, Mackie2002}, $^{e}$\citet{Plane2021}, $^{f}$\citet{Nava1989}, $^{g}$\citet{Fritz1982}, $^{h}$\citet{Baptista2023}, $^{i}$This study, $^{j}$\citet{Lizardo-Huerta2021}.

The coefficients follow the VULCAN formatting and units \citep{Tsai2017, Tsai2021}, where the generalised Arrhenius equation is used
\begin{equation}
    k = A T^{b} \exp(-\frac{E}{T}),
\end{equation}
where $k$ is the rate coefficient in units of cm$^{3}$s$^{-1}$ for bimolecular reactions and cm$^{6}$ s$^{-1}$ for termolecular reactions.
We follow convention where odd numbers are the forward reactions and even numbers the reverse reactions.

\startlongtable
\begin{deluxetable}{llccc} 
\tablehead{
\colhead{Reaction number} & \colhead{Forward reaction (2-body)} & \dcolhead{A} & \dcolhead{n} & \dcolhead{E}}
\startdata
R1 & O + OH $\rightarrow$ \ce{O2} + H & 7.47e-10 & -0.5 & 30.0\\
R3 & OH + \ce{H2} $\rightarrow$ \ce{H2O} + H & 3.57e-16 & 1.52 & 1740.0\\
R5 & O + \ce{H2O} $\rightarrow$ OH + OH & 8.2e-14 & 0.95 & 8570.0\\
R7 & O + \ce{H2} $\rightarrow$ OH + H & 8.52e-20 & 2.67 & 3160.0\\
R9 & O$_{1}$ + \ce{H2} $\rightarrow$ OH + H & 2.87e-10 & 0.0 & 0.0\\
R11 & O$_{1}$ + \ce{O2} $\rightarrow$ O + \ce{O2} & 3.2e-11 & 0.0 & -70.0\\
R13 & O$_{1}$ + \ce{H2O} $\rightarrow$ OH + OH & 1.62e-10 & 0.0 & -65.0\\
R15 & \ce{HO2} + H $\rightarrow$ OH + OH & 2.81e-10 & 0.0 & 440.0\\
R17 & \ce{HO2} + H $\rightarrow$ \ce{O2} + \ce{H2} & 7.11e-11 & 0.0 & 710.0\\
R19 & O + \ce{HO2} $\rightarrow$ OH + \ce{O2} & 2.7e-11 & 0.0 & -224.0\\
R21 & OH + \ce{HO2} $\rightarrow$ \ce{H2O} + \ce{O2} & 2.4e-08 & -1.0 & 0.0\\
R23 & \ce{H2O2} + H $\rightarrow$ \ce{H2} + \ce{HO2} & 2.81e-12 & 0.0 & 1890.0\\
R25 & \ce{H2O2} + H $\rightarrow$ OH + \ce{H2O} & 1.69e-11 & 0.0 & 1800.0\\
R27 & O + \ce{H2O2} $\rightarrow$ OH + \ce{HO2} & 1.4e-12 & 0.0 & 2000.0\\
R29 & OH + \ce{H2O2} $\rightarrow$ \ce{H2O} + \ce{HO2} & 2.9e-12 & 0.0 & 160.0\\
R31$^{a}$ & OH + \ce{HOPO2} $\rightarrow$ \ce{PO3} + \ce{H2O} & 1.993e-18 & 2.0 & 1007.0\\
R33$^{a}$ & \ce{H2} + \ce{PO3} $\rightarrow$ H + \ce{HOPO2} & 3.321e-12 & 0.0 & 0.0\\
R35$^{b}$ & \ce{O2} + PO $\rightarrow$ O + \ce{PO2} & 2.3e-11 & 0.0 & 100.0\\
R37$^{b}$ & \ce{O2} + P $\rightarrow$ O + PO & 4.2e-12 & 0.0 & 600.0\\
R39$^{c}$ & \ce{O2} + PH $\rightarrow$ O + HPO & 5.25e-13 & 0.0 & 2012.16\\
R41$^{a}$ & H + HOPO $\rightarrow$ \ce{H2O} + PO & 4.98e-12 & 0.0 & 4176.72\\
R43$^{d}$ & H + HOPO $\rightarrow$ \ce{H2} + \ce{PO2} & 3.55e-17 & 1.94 & 5072.45\\
R45$^{d}$ & H + \ce{HOPO2} $\rightarrow$ \ce{H2O} + \ce{PO2} & 1.78e-11 & 0.176 & 5937.99\\
R47$^{a}$ & O + HOPO $\rightarrow$ H + \ce{PO3} & 1.66e-12 & 0.0 & 7548.29\\
R49$^{e}$ & H + \ce{PO3} $\rightarrow$ OH + \ce{PO2} & 1.16e-11 & 0.5 & 0.0\\
R51$^{c}$ & H + \ce{P2O3} $\rightarrow$ PO + HOPO & 5.25e-11 & 0.0 & 6013.62\\
R53$^{a}$ & H + HPO $\rightarrow$ \ce{H2} + PO & 4e-16 & 1.5 & 0.0\\
R55$^{i}$ & H + \ce{PH3} $\rightarrow$ \ce{H2} + \ce{PH2} & 7e-18 & 2.3576 & 45.123\\
R57$^{i}$ & H + \ce{PH2} $\rightarrow$ \ce{H2} + PH & 1.94e-16 & 1.8025 & 47.26\\
R59$^{i}$ & H + PH $\rightarrow$ P + \ce{H2} & 1.54e-15 & 1.5073 & 5.7185\\
R61$^{c}$ & H + \ce{P2O} $\rightarrow$ OH + \ce{P2} & 5.25e-11 & 0.0 & 2807.16\\
R63$^{c}$ & H + \ce{P2O} $\rightarrow$ PO + PH & 5.25e-11 & 0.0 & 2810.76\\
R65$^{c}$ & H + \ce{P2O} $\rightarrow$ HPO + P & 5.25e-11 & 0.0 & 6013.62\\
R67$^{c}$ & H + \ce{P2O2} $\rightarrow$ PO + HPO & 5.25e-11 & 0.0 & 6013.62\\
R69$^{c}$ & H + \ce{H2POH} $\rightarrow$ \ce{H2O} + \ce{PH2} & 5.25e-11 & 0.0 & 6013.62\\
R71$^{c}$ & H + \ce{H2POH} $\rightarrow$ \ce{H2} + HPOH & 5.25e-11 & 0.0 & 2089.13\\
R73$^{c}$ & H + HPOH $\rightarrow$ \ce{H2O} + PH & 5.25e-11 & 0.0 & 0.0\\
R75$^{c}$ & H + HPOH $\rightarrow$ \ce{H2} + HPO & 5.25e-11 & 0.0 & 2863.68\\
R77$^{a}$ & O + HOPO $\rightarrow$ OH + \ce{PO2} & 1.66e-11 & 0.0 & 0.0\\
R79$^{c}$ & O + \ce{HOPO2} $\rightarrow$ \ce{O2} + HOPO & 5.25e-11 & 0.0 & 4150.6\\
R81$^{e}$ & O + \ce{PO3} $\rightarrow$ \ce{O2} + \ce{PO2} & 5.04e-11 & -0.04 & 0.0\\
R83$^{c}$ & O + \ce{P2O3} $\rightarrow$ PO + \ce{PO3} & 5.25e-11 & 0.0 & 6013.62\\
R85$^{c}$ & O + \ce{P2O3} $\rightarrow$ \ce{PO2} + \ce{PO2} & 5.25e-11 & 0.0 & 6013.62\\
R87$^{a}$ & O + HPO $\rightarrow$ H + \ce{PO2} & 1.66e-11 & 0.0 & 1511.0\\
R89$^{a}$ & O + HPO $\rightarrow$ OH + PO & 2.823e-16 & 1.5 & 0.0\\
R91$^{c}$ & O + \ce{P2} $\rightarrow$ PO + P & 5.25e-11 & 0.0 & 2288.78\\
R93$^{c}$ & O + \ce{PH3} $\rightarrow$ OH + \ce{PH2} & 2.855e-18 & 2.296 & 915.6\\
R95$^{f}$ & O + \ce{PH3} $\rightarrow$ HPOH + H & 4.75e-11 & 0.0 & 0.0\\
R97$^{c}$ & O + \ce{PH2} $\rightarrow$ H + HPO & 5.25e-11 & 0.0 & 0.0\\
R99$^{c}$ & O + \ce{PH2} $\rightarrow$ OH + PH & 5.25e-11 & 0.0 & 1864.22\\
R101$^{b}$ & O + PH $\rightarrow$ PO + H & 2e-10 & 0.0 & 0.0\\
R103$^{c}$ & O + PH $\rightarrow$ OH + P & 5.25e-11 & 0.0 & 1873.84\\
R105$^{c}$ & O + \ce{P2O} $\rightarrow$ \ce{O2} + \ce{P2} & 5.25e-11 & 0.0 & 1704.26\\
R107$^{c}$ & O + \ce{P2O} $\rightarrow$ PO + PO & 5.25e-11 & 0.0 & 849.12\\
R109$^{c}$ & O + \ce{P2O} $\rightarrow$ \ce{PO2} + P & 5.25e-11 & 0.0 & 6013.62\\
R111$^{c}$ & O + \ce{P2O2} $\rightarrow$ \ce{O2} + \ce{P2O} & 5.25e-11 & 0.0 & 3089.8\\
R113$^{c}$ & O + \ce{P2O2} $\rightarrow$ PO + \ce{PO2} & 5.25e-11 & 0.0 & 6013.62\\
R115$^{c}$ & O + \ce{H2POH} $\rightarrow$ OH + HPOH & 5.25e-11 & 0.0 & 1408.39\\
R117$^{c}$ & O + HPOH $\rightarrow$ H + HOPO & 5.25e-11 & 0.0 & 0.0\\
R119$^{c}$ & O + HPOH $\rightarrow$ OH + HPO & 5.25e-11 & 0.0 & 2310.43\\
R121$^{b}$ & OH + PO $\rightarrow$ H + \ce{PO2} & 1.2e-10 & 0.0 & 0.0\\
R123$^{d}$ & OH + HOPO $\rightarrow$ \ce{H2O} + \ce{PO2} & 6.17e-11 & -0.219 & 1610.3\\
R125$^{e}$ & OH + HOPO $\rightarrow$ H + \ce{HOPO2} & 7.69e-08 & -1.25 & 0.0\\
R127$^{a}$ & O + \ce{HOPO2} $\rightarrow$ OH + \ce{PO3} & 1.66e-11 & 0.0 & 6194.0\\
R129$^{c}$ & OH + \ce{P2O3} $\rightarrow$ PO + \ce{HOPO2} & 5.25e-13 & 0.0 & 6013.62\\
R131$^{c}$ & OH + \ce{P2O3} $\rightarrow$ \ce{PO2} + HOPO & 5.25e-13 & 0.0 & 6013.62\\
R133$^{a}$ & OH + HPO $\rightarrow$ \ce{H2O} + PO & 2e-18 & 2.0 & 1007.0\\
R135$^{c}$ & OH + HPO $\rightarrow$ H + HOPO & 5.25e-13 & 0.0 & 6013.62\\
R137$^{b}$ & OH + P $\rightarrow$ H + PO & 3.61e-11 & -0.29 & 0.0\\
R139$^{g}$ & OH + \ce{PH3} $\rightarrow$ \ce{H2O} + \ce{PH2} & 2.71e-11 & 0.0 & 155.15\\
R141$^{c}$ & OH + \ce{PH3} $\rightarrow$ H + \ce{H2POH} & 5.25e-13 & 0.0 & 6013.62\\
R143$^{c}$ & OH + \ce{PH2} $\rightarrow$ \ce{H2O} + PH & 5.25e-13 & 0.0 & 1126.95\\
R145$^{c}$ & OH + \ce{PH2} $\rightarrow$ H + HPOH & 5.25e-13 & 0.0 & 0.0\\
R147$^{b}$ & OH + PH $\rightarrow$ \ce{H2O} + P & 3.86e-11 & 0.167 & 0.0\\
R149$^{c}$ & OH + PH $\rightarrow$ H + HPO & 5.25e-13 & 0.0 & 2287.58\\
R151$^{c}$ & OH + \ce{P2O} $\rightarrow$ H + \ce{P2O2} & 5.25e-13 & 0.0 & 7167.03\\
R153$^{c}$ & OH + \ce{P2O} $\rightarrow$ HOPO + P & 5.25e-13 & 0.0 & 6013.62\\
R155$^{c}$ & OH + \ce{P2O2} $\rightarrow$ PO + HOPO & 5.25e-13 & 0.0 & 6013.62\\
R157$^{c}$ & OH + \ce{H2POH} $\rightarrow$ \ce{H2O} + HPOH & 5.25e-13 & 0.0 & 823.87\\
R159$^{c}$ & OH + HPOH $\rightarrow$ \ce{H2O} + HPO & 5.25e-13 & 0.0 & 1650.14\\
R161$^{c}$ & \ce{HO2} + PO $\rightarrow$ \ce{O2} + HPO & 5.25e-13 & 0.0 & 3396.49\\
R163$^{c}$ & \ce{HO2} + PO $\rightarrow$ O + HOPO & 5.25e-13 & 0.0 & 0.0\\
R165$^{a}$ & \ce{HO2} + PO $\rightarrow$ OH + \ce{PO2} & 3.49e-12 & 0.0 & -251.61\\
R167$^{a}$ & \ce{O2} + HOPO $\rightarrow$ \ce{HO2} + \ce{PO2} & 1.16e-11 & 0.0 & 22795.84\\
R169$^{c}$ & \ce{HO2} + \ce{PO2} $\rightarrow$ O + \ce{HOPO2} & 5.25e-13 & 0.0 & 0.0\\
R171$^{a}$ & \ce{HO2} + \ce{PO2} $\rightarrow$ OH + \ce{PO3} & 8.3e-13 & 0.0 & 0.0\\
R173$^{a}$ & \ce{HO2} + HOPO $\rightarrow$ OH + \ce{HOPO2} & 2.49e-10 & 0.0 & 11875.98\\
R175$^{a}$ & \ce{HO2} + HOPO $\rightarrow$ \ce{H2O2} + \ce{PO2} & 4.15e-12 & 0.0 & 11725.02\\
R177$^{a}$ & \ce{HO2} + \ce{HOPO2} $\rightarrow$ \ce{H2O2} + \ce{PO3} & 4.15e-12 & 0.0 & 12379.2\\
R179$^{a}$ & \ce{O2} + \ce{HOPO2} $\rightarrow$ \ce{HO2} + \ce{PO3} & 1.16e-11 & 0.0 & 33212.49\\
R181$^{c}$ & \ce{HO2} + HPO $\rightarrow$ \ce{O2} + HPOH & 5.25e-13 & 0.0 & 5135.63\\
R183$^{c}$ & \ce{HO2} + P $\rightarrow$ \ce{O2} + PH & 5.25e-11 & 0.0 & 3420.55\\
R185$^{c}$ & \ce{HO2} + P $\rightarrow$ OH + PO & 5.25e-11 & 0.0 & 922.49\\
R187$^{c}$ & \ce{HO2} + \ce{P2} $\rightarrow$ OH + \ce{P2O} & 5.25e-13 & 0.0 & 2549.77\\
R189$^{c}$ & \ce{HO2} + \ce{PH2} $\rightarrow$ \ce{O2} + \ce{PH3} & 5.25e-13 & 0.0 & 2639.98\\
R191$^{c}$ & \ce{HO2} + \ce{PH2} $\rightarrow$ O + \ce{H2POH} & 5.25e-13 & 0.0 & 0.0\\
R193$^{c}$ & \ce{HO2} + PH $\rightarrow$ \ce{O2} + \ce{PH2} & 5.25e-13 & 0.0 & 2629.15\\
R195$^{c}$ & \ce{HO2} + PH $\rightarrow$ O + HPOH & 5.25e-13 & 0.0 & 0.0\\
R197$^{c}$ & \ce{HO2} + PH $\rightarrow$ OH + HPO & 5.25e-13 & 0.0 & 796.2\\
R199$^{c}$ & \ce{HO2} + \ce{P2O} $\rightarrow$ OH + \ce{P2O2} & 5.25e-13 & 0.0 & 2265.93\\
R201$^{c}$ & \ce{HO2} + HPOH $\rightarrow$ \ce{O2} + \ce{H2POH} & 5.25e-13 & 0.0 & 2658.02\\
R203$^{c}$ & PO + \ce{HOPO2} $\rightarrow$ \ce{PO2} + HOPO & 5.25e-13 & 0.0 & 4895.08\\
R205$^{a}$ & PO + \ce{PO3} $\rightarrow$ \ce{PO2} + \ce{PO2} & 8.3e-13 & 0.0 & 0.0\\
R207$^{c}$ & PO + \ce{P2O} $\rightarrow$ \ce{PO2} + \ce{P2} & 5.25e-13 & 0.0 & 2050.64\\
R209$^{c}$ & PO + \ce{P2O2} $\rightarrow$ \ce{PO2} + \ce{P2O} & 5.25e-13 & 0.0 & 3612.98\\
R211$^{c}$ & PO + \ce{H2POH} $\rightarrow$ HOPO + \ce{PH2} & 5.25e-13 & 0.0 & 6013.62\\
R213$^{c}$ & PO + HPOH $\rightarrow$ HOPO + PH & 5.25e-13 & 0.0 & 0.0\\
R215$^{c}$ & PO + HPOH $\rightarrow$ HPO + HPO & 5.25e-13 & 0.0 & 5082.71\\
R217$^{c}$ & \ce{PO2} + HPO $\rightarrow$ H + \ce{P2O3} & 5.25e-13 & 0.0 & 6013.62\\
R219$^{a}$ & \ce{PO2} + HPO $\rightarrow$ PO + HOPO & 3.321e-13 & 0.0 & 0.0\\
R221$^{c}$ & \ce{PO2} + P $\rightarrow$ PO + PO & 5.25e-11 & 0.0 & 2472.8\\
R223$^{c}$ & \ce{PO2} + \ce{PH3} $\rightarrow$ HOPO + \ce{PH2} & 5.25e-13 & 0.0 & 0.0\\
R225$^{c}$ & \ce{PO2} + \ce{PH2} $\rightarrow$ HOPO + PH & 5.25e-13 & 0.0 & 0.0\\
R227$^{c}$ & \ce{PO2} + PH $\rightarrow$ PO + HPO & 5.25e-13 & 0.0 & 2418.68\\
R229$^{c}$ & \ce{PO2} + PH $\rightarrow$ HOPO + P & 5.25e-13 & 0.0 & 36.08\\
R231$^{c}$ & \ce{PO2} + \ce{P2O} $\rightarrow$ \ce{PO3} + \ce{P2} & 5.25e-13 & 0.0 & 309.1\\
R233$^{c}$ & \ce{PO2} + \ce{P2O} $\rightarrow$ \ce{P2O3} + P & 5.25e-13 & 0.0 & 6013.62\\
R235$^{c}$ & \ce{PO2} + \ce{P2O2} $\rightarrow$ PO + \ce{P2O3} & 5.25e-13 & 0.0 & 6013.62\\
R237$^{c}$ & \ce{PO2} + \ce{H2POH} $\rightarrow$ HOPO + HPOH & 5.25e-13 & 0.0 & 0.0\\
R239$^{c}$ & \ce{PO2} + \ce{H2POH} $\rightarrow$ \ce{HOPO2} + \ce{PH2} & 5.25e-13 & 0.0 & 6013.62\\
R241$^{c}$ & \ce{PO2} + HPOH $\rightarrow$ HOPO + HPO & 5.25e-13 & 0.0 & 894.83\\
R243$^{c}$ & \ce{PO2} + HPOH $\rightarrow$ \ce{HOPO2} + PH & 5.25e-13 & 0.0 & 0.0\\
R245$^{a}$ & HOPO + \ce{PO3} $\rightarrow$ \ce{PO2} + \ce{HOPO2} & 8.3e-13 & 0.62 & 0.0\\
R247$^{c}$ & HOPO + \ce{P2O} $\rightarrow$ \ce{HOPO2} + \ce{P2} & 5.25e-13 & 0.0 & 18040.85\\
R249$^{c}$ & HOPO + \ce{P2O2} $\rightarrow$ \ce{HOPO2} + \ce{P2O} & 5.25e-13 & 0.0 & 18040.85\\
R251$^{c}$ & \ce{HOPO2} + P $\rightarrow$ PO + HOPO & 5.25e-11 & 0.0 & 3445.8\\
R253$^{c}$ & \ce{HOPO2} + PH $\rightarrow$ HOPO + HPO & 5.25e-13 & 0.0 & 3502.33\\
R255$^{a}$ & \ce{PO3} + HPO $\rightarrow$ PO + \ce{HOPO2} & 3.321e-13 & 0.0 & 0.0\\
R257$^{c}$ & \ce{PO3} + P $\rightarrow$ PO + \ce{PO2} & 5.25e-11 & 0.0 & 18.04\\
R259$^{c}$ & \ce{PO3} + \ce{PH3} $\rightarrow$ \ce{HOPO2} + \ce{PH2} & 5.25e-13 & 0.0 & 0.0\\
R261$^{c}$ & \ce{PO3} + \ce{PH2} $\rightarrow$ \ce{HOPO2} + PH & 5.25e-13 & 0.0 & 0.0\\
R263$^{c}$ & \ce{PO3} + PH $\rightarrow$ \ce{PO2} + HPO & 5.25e-13 & 0.0 & 0.0\\
R265$^{c}$ & \ce{PO3} + PH $\rightarrow$ \ce{HOPO2} + P & 5.25e-13 & 0.0 & 0.0\\
R267$^{c}$ & \ce{PO3} + \ce{P2O} $\rightarrow$ \ce{PO2} + \ce{P2O2} & 5.25e-13 & 0.0 & 917.68\\
R269$^{c}$ & \ce{PO3} + \ce{H2POH} $\rightarrow$ \ce{HOPO2} + HPOH & 5.25e-13 & 0.0 & 0.0\\
R271$^{c}$ & \ce{PO3} + HPOH $\rightarrow$ \ce{HOPO2} + HPO & 5.25e-13 & 0.0 & 0.0\\
R273$^{c}$ & HPO + P $\rightarrow$ PO + PH & 5.25e-11 & 0.0 & 3474.67\\
R275$^{c}$ & HPO + \ce{PH2} $\rightarrow$ PO + \ce{PH3} & 5.25e-13 & 0.0 & 2466.79\\
R277$^{c}$ & HPO + PH $\rightarrow$ PO + \ce{PH2} & 5.25e-13 & 0.0 & 2475.21\\
R279$^{c}$ & HPO + HPOH $\rightarrow$ PO + \ce{H2POH} & 5.25e-13 & 0.0 & 2489.64\\
R281$^{i}$ & P + PH $\rightarrow$ H + \ce{P2} & 3.66e-11 & 0.198 & -1.166\\
R283$^{c}$ & P + \ce{P2O} $\rightarrow$ PO + \ce{P2} & 5.25e-11 & 0.0 & 1467.32\\
R285$^{c}$ & P + \ce{P2O2} $\rightarrow$ PO + \ce{P2O} & 5.25e-11 & 0.0 & 2815.58\\
R287$^{c}$ & P + HPOH $\rightarrow$ HPO + PH & 5.25e-11 & 0.0 & 4874.64\\
R289$^{c}$ & \ce{PH3} + PH $\rightarrow$ \ce{PH2} + \ce{PH2} & 5.25e-13 & 0.0 & 2759.05\\
R291$^{c}$ & \ce{PH3} + HPOH $\rightarrow$ \ce{PH2} + \ce{H2POH} & 5.25e-13 & 0.0 & 2830.01\\
R293$^{i}$ & \ce{PH2} + PH $\rightarrow$ P + \ce{PH3} & 2.51e-21 & 2.9224 & -240.52\\
R295$^{c}$ & \ce{PH2} + HPOH $\rightarrow$ HPO + \ce{PH3} & 5.25e-13 & 0.0 & 3825.86\\
R297$^{c}$ & PH + PH $\rightarrow$ P + \ce{PH2} & 5.25e-13 & 0.0 & 3317.11\\
R299$^{c}$ & PH + \ce{P2O} $\rightarrow$ HPO + \ce{P2} & 5.25e-13 & 0.0 & 1721.1\\
R301$^{c}$ & PH + \ce{P2O2} $\rightarrow$ HPO + \ce{P2O} & 5.25e-13 & 0.0 & 2804.75\\
R303$^{c}$ & PH + \ce{H2POH} $\rightarrow$ \ce{PH2} + HPOH & 5.25e-13 & 0.0 & 2731.39\\
R305$^{c}$ & PH + HPOH $\rightarrow$ HPO + \ce{PH2} & 5.25e-13 & 0.0 & 3675.52\\
R307$^{c}$ & PH + HPOH $\rightarrow$ P + \ce{H2POH} & 5.25e-13 & 0.0 & 3479.48\\
R309$^{c}$ & \ce{P2O} + \ce{P2O} $\rightarrow$ \ce{P2} + \ce{P2O2} & 5.25e-13 & 0.0 & 18040.85\\
R311$^{c}$ & HPOH + HPOH $\rightarrow$ HPO + \ce{H2POH} & 5.25e-13 & 0.0 & 3858.34\\
R313$^{j}$ & \ce{H3PO4} $\rightarrow$ \ce{HOPO2} + \ce{H2O} & 8.81e4 & 2.12 & 19604.37 \\
R315 & He $\rightarrow$ He & 0.0 & 0.0 & 0.0\\
\enddata
\end{deluxetable}

\startlongtable
\begin{deluxetable}{llcccccc} 
\tablehead{
\colhead{Reaction number} & \colhead{Forward reaction (3-body)} & \dcolhead{A} & \dcolhead{n} & \dcolhead{E} & \dcolhead{A_{\infty}} & \dcolhead{n_{\infty}} & \dcolhead{E_{\infty}}}
\startdata
R317 & H + H + M $\rightarrow$ \ce{H2} + M & 2.7e-31 & -0.6 & 0.0 & 3.31e-06 & -1.0 & 0.0\\
R319 & H + O + M $\rightarrow$ OH + M & 1.3e-29 & -1.0 & 0.0 & 1e-11 & 0.0 & 0.0\\
R321 & OH + H + M $\rightarrow$ \ce{H2O} + M & 3.89e-25 & -2.0 & 0.0 & 4.26e-11 & 0.23 & -57.5\\
R323 & H + \ce{O2} + M $\rightarrow$ \ce{HO2} + M & 2.17e-29 & -1.1 & 0.0 & 7.51e-11 & 0.0 & 0.0\\
R325 & \ce{HO2} + \ce{HO2} + M $\rightarrow$ \ce{H2O2} + \ce{O2} + M & 1.9e-33 & 0.0 & -980.0 & 2.2e-13 & 0.0 & -600.0\\
R327 & OH + OH + M $\rightarrow$ \ce{H2O2} + M & 7.97e-31 & -0.76 & 0.0 & 1.51e-11 & -0.37 & 0.0\\
R329$^{i}$ & H + \ce{PH2} + M $\rightarrow$ \ce{PH3} + M & 4.320e-24 &  -2.1662 & 211.18 &  1.220e-10  &  0.200  &    -8.013\\
R331$^{d}$ & H + \ce{PO2} + M $\rightarrow$ HOPO + M & 7.95e-17 & -4.33 & 513.28 & 1.91e-14 & 1.29 & -754.83\\
R333$^{d}$ & OH + \ce{PO2} + M $\rightarrow$ \ce{HOPO2} + M & 0.28 & -8.59 & 4528.98 & 2.57e-10 & -0.24 & 0.0\\
R335$^{i}$ & H + \ce{P2} + M $\rightarrow$ P2H + M & 2.47e-27 & -1.23 & 152.0 & 1.45e-11 & 0.54 & -58.9\\
R337$^{i}$ & PO + \ce{PO2} + M $\rightarrow$ \ce{P2O3} + M & 1.49e-15 & -4.25 & 0.0 & 2.819e-10 & 0.167 & 0.0\\
R339$^{c,h}$ & H + P + M $\rightarrow$ PH + M & 9.26e-30 & -1.1 & 357.21 & 1.79e-12 & -0.13 & 459.99\\
R341$^{i}$ & H + PH + M $\rightarrow$ \ce{PH2} + M & 2.91e-28 &  -1.10     94.5  & 7.98e-11   &  0.222   &  0.535\\
R343$^{i}$ & \ce{PH2} + \ce{PH2} + M $\rightarrow$ \ce{P2H4} + M & 5.48e-15 & -4.836 & 351.6 & 1.67e-10 & -0.105 & 45.0\\
R345$^{i}$ & \ce{P2H2} + \ce{H2} + M $\rightarrow$ \ce{P2H4} + M & 6.55e-10 & -6.99 & 6363.0 & 8.51e-19 & 2.238 & 4674.0\\
R347$^{c,h}$ & OH + P + M $\rightarrow$ HPO + M & 1.241e-25 & -1.95 & 670.0 & 4.12e-10 & 0.16 & 128.41\\
R349$^{i}$ & \ce{P2O3} + \ce{P2O3} + M $\rightarrow$ P4O6 + M & 5.925e-18 & -2.99 & 0.0 & 3.263e-10 & 0.166 & 0.0\\
R351 & O + O + M $\rightarrow$ \ce{O2} + M & 5.21e-35 & 0.0 & -900.0 & - & - & - \\
R353$^{a}$ & H + PO + M $\rightarrow$ HPO + M & 1.241e-25 & -1.95 & 670.0 & - & - & -\\
R355$^{a}$ & H + \ce{PO3} + M $\rightarrow$ \ce{HOPO2} + M & 3.309e-23 & -2.37 & 720.0 & - & - & - \\
R357$^{c}$ & H + HPO + M $\rightarrow$ HPOH + M & 7.549e-26 & -1.422 & 415.5 & - & - & - \\
R359$^{c}$ & H + HPOH + M $\rightarrow$ \ce{H2POH} + M & 9.619e-24 & -1.885 & 550.8 & - & - & - \\
R361$^{a}$ & O + PO + M $\rightarrow$ \ce{PO2} + M & 1.103e-22 & -2.63 & 866.0 & - & - & - \\
R363$^{a}$ & O + \ce{PO2} + M $\rightarrow$ \ce{PO3} + M & 8.962e-21 & -3.15 & 946.7 & - & - & - \\
R365$^{a}$ & O + HOPO + M $\rightarrow$ \ce{HOPO2} + M & 8.273e-21 & -2.99 & 1027.0 & - & - & - \\
R367$^{c}$ & O + P + M $\rightarrow$ PO + M & 1.642e-29 & -0.747 & 218.2 & - & - & - \\
R369$^{a}$ & OH + PO + M $\rightarrow$ HOPO + M & 6.894e-27 & -2.09 & 800.7 & - & - & - \\
R371$^{c}$ & OH + \ce{PH2} + M $\rightarrow$ \ce{H2POH} + M & 5.715e-29 & -1.223 & 357.2 & - & - & - \\
R373$^{c}$ & OH + PH + M $\rightarrow$ HPOH + M & 2.175e-33 & -0.415 & 121.4 & - & - & - \\
R375$^{c}$ & O + \ce{P2} + M $\rightarrow$ \ce{P2O} + M & 7.774e-31 & -0.844 & 265.5 & - & - & - \\
R377$^{c}$ & O + PH + M $\rightarrow$ HPO + M & 2.162e-33 & -0.309 & 97.2 & - & - & - \\
R379$^{c}$ & O + \ce{P2O} + M $\rightarrow$ \ce{P2O2} + M & 5.995e-34 & -0.268 & 84.4 & - & - & - \\
R381$^{c}$ & PO + PO + M $\rightarrow$ \ce{P2O2} + M & 2.117e-28 & -2.077 & 595.2 & - & - & - \\
R383$^{c}$ & PO + P + M $\rightarrow$ \ce{P2O} + M & 2.64e-24 & -2.41 & 690.7 & - & - & - \\
R385$^{b}$ & \ce{HOPO2} + \ce{H2O} + M $\rightarrow$  \ce{H3PO4} + M & 1.35e-07 & -7.53 & 0.0 & - & - & - \\
R387$^{c}$ & P + P + M $\rightarrow$ \ce{P2} + M & 7.191e-27 & -1.67 & 477.2 & - & - & - \\
R389$^{c}$ & \ce{P2} + \ce{P2} + M $\rightarrow$ P4 + M & 3.721e-26 & -1.867 & 545.4 & - & - & - \\
\enddata
\end{deluxetable}

\section{Wang et al. (2017) comparison}
\label{app:wang}

\end{document}